\documentclass[a4paper]{article}

\usepackage{graphicx}
\usepackage{color}
\usepackage{placeins}
\usepackage{float}
\usepackage{tabularx,colortbl}

\usepackage{times,amsmath,epsfig,algorithm,algorithmic,comment,amssymb, bm}
\usepackage{subfigure}

\usepackage{etoolbox}
\newtoggle{showrevisions}

\toggletrue{showrevisions}

\iftoggle{showrevisions}{%
\usepackage{xcolor}
\usepackage{changebar}
\usepackage[normalem]{ulem}

\newcommand{\removed}[1]{\cbstart\removedfragile{#1}\cbend{}}
\newcommand{\removedfragile}[1]{{\color{red}{\sout{#1}}}{}}

}{%
  \newcommand{\removed}[1]{} 
  \newcommand{\removedfragile}[1]{}

}

\newcommand{\R}{\mathbb{R}}
\newcommand{\N}{\mathbb{N}}

\DeclareMathOperator*{\argmin}{arg\,min}

\newcommand{\gnuplotFigResizeFactor}{0.40}

\begin{document}

\title{GPU-Accelerated Algorithms for\\ Compressed Signals Recovery with Application to Astronomical Imagery Deblurring}
\author{Attilio Fiandrotti, Sophie M. Fosson, Chiara Ravazzi, and Enrico Magli\\
Dipartimento di Elettronica e Telecomunicazioni\\
Politecnico di Torino, Italy\\
attilio.fiandrotti@polito.it, sophie.fosson@polito.it,\\ chiara.ravazzi@polito.it, enrico.magli@polito.it\footnote{This work was supported by the European Research Council under FP7 / ERC, Grant agreement n.279848 - CRISP project.}}

\maketitle

\begin{abstract}
Compressive sensing promises to enable bandwidth-efficient on-board compression of astronomical data by lifting the encoding complexity from the source to the receiver.
The signal is recovered off-line, exploiting GPUs parallel computation capabilities to speedup the reconstruction process.
However, inherent GPU hardware constraints limit the size of the recoverable signal and the speedup practically achievable.
In this work, we design parallel algorithms that exploit the properties of circulant matrices for efficient GPU-accelerated sparse signals recovery.
Our approach reduces the memory requirements, allowing us to recover very large signals with limited memory.
In addition, it achieves a tenfold signal recovery speedup thanks to ad-hoc parallelization of matrix-vector multiplications and matrix inversions.
Finally, we practically demonstrate our algorithms in a typical application of circulant matrices: deblurring a sparse astronomical image in the compressed domain.
\end{abstract}


\section{Introduction}
\label{sec:introduction}

Compressive Sensing (CS, \cite{candes08}) has drawn a lot of interest for a number of remote sensing applications (see \cite{Bob2008} and the references therein).
The CS paradigm is as follows. Let $\bm{x}^{\star}=(x^{\star}_1,\dots, x^{\star}_n)$ be the \emph{sparse} signal to be sensed, \emph{i.e.}, at most $k \ll n$ elements of $\bm{x}^{\star}$ are different from zero.
Let $\mathbf{A}$ be the $m \times n$ \emph{sensing matrix} ($m < n$): the sensing process is expressed by the linear combination

\begin{equation}\label{model}
\bm{y}=\mathbf{A}\bm{x}^{\star}
\end{equation}
\noindent
where $\bm{y} = (y_\textit{1}, ..., y_\textit{m})$ is the \emph{measurements vector}.
Under certain conditions, CS theory shows that  $\bm{x}^{\star}$ can be exactly recovered seeking the sparsest solution that fulfills $\mathbf{A} \bm{x}^{\star} = \bm{y}$ despite $m < n$ \cite{can06, can08}.
In particular, CS low-complexity coding process makes it well suited for power and bandwidth-constrained applications such as spaceborne data compression \cite{4776452}.

While CS significantly reduces the onboard sensing process complexity, such complexity is lifted to the receiver, which is tasked with recovering the signal.
Finding the sparsest solution $x$ fulfilling $\bm{y} = \mathbf{A}\bm{x}$ is an NP-hard optimization problem \cite[Section 2.3]{fou13}, so approximate solutions are practically sought using the $\ell_1$-norm.
Namely, such problems are typically recast in a convex form, e.g., LASSO \cite{tib94}, that can be solved by Iterative Soft Thresholding (ISTA \cite{dau04, for10}) or Alternating Direction Methods of Multipliers (ADMM, \cite{boy10}). Both ISTA and ADMM have been proved to converge linearly to a LASSO minimum, however ADMM requires fewer iterations to converge at the expense of slightly more complex update step and larger memory requirements (preliminarily inverting and storing a $n\times n$ matrix is required). In both cases, recovery time increases with $n$, which is an issue with very large signals.

In the last years, the use of Graphical Processing Units (GPUs)  to speedup the recovery of compressively sensed signals has attracted increasing attention.
A GPU is a multicore processor whose cores operate in parallel on the data stored in some dedicated GPU memory. \cite{blanchard2012gpu} proposed greedy algorithms to solve an $\ell_0$ formulation of the CS problem, which is however an NP-complex problem. On the contrary, the $\ell_1$ problem formulation (as in LASSO) is practically tractable in reason of its convexity. Moreover, LASSO-solving algorithms such as ISTA and ADMM are well-suited for GPU acceleration as their complexity lies in inherently parallel matrix-vector multiplications. For example \cite{tian2012high} experimented GPU-based radar signals recovery using an IST algorithm. In \cite{ber16}, GPUs are exploited to accelerate the recovery in hyperspectral imaging with HYCA method (which relies on ADMM).  In \cite{mur12, qua15}, multi-GPU systems are leveraged for CS in magnetic resonance imaging (MRI).


The performance of GPU-accelerated algorithms is however hindered by some inherent hardware constraints.
First and foremost, the amount of available GPU memory limits the maximum size of the recoverable signal.
The memory footprint of data structures such as the sensing matrix grows in fact with the signal size $n$, so large signals may be intractable on GPUs.
Second, a GPU effective degree of parallelism is upper-bounded by the bandwidth available on the bus towards the GPU memory.
In fact, as data structures become larger, memory caches become ineffective and the GPU cores line up for accessing the memory over a shared bus.
As large signals are increasingly commonplace in remote sensing applications, being able to practically recover such signals within the CS paradigm is becoming a relevant issue.

In this work, we leverage circulant matrices to efficiently recover large compressively sensed signals on GPUs \cite{yin10, ra10}.
A circulant matrix is such that each row is obtained as the circular shift of the previous row.
For the problem of remotely sensing large signals, circulant matrices are attractive because (a) they are suitable sensing matrices for CS \cite{ra10}; (b) their memory footprint grows only linearly with $n$, fitting in low-complexity spaceborne encoders (c); they allow to handle otherwise complex inversion tasks with FFT at signal recovery time, as explained in the following.

The pivotal contributions of this work are CPADMM and CPISTA, two GPU-accelerated LASSO-solving algorithms for sparse signals recovery.
Both algorithms leverage circulant sensing matrices properties to achieve a memory-efficient sensing matrix representation.
Namely, CPISTA improves over the PISTA algorithm in \cite{pcs2013} by exploiting circulant matrices properties to speedup matrix-vector multiplications. 
CPADMM enjoys all the benefits of CPISTA, plus it addresses ADMM inversion complexity issue decomposing the sensing matrix as a product of a fat projection matrix and a square circulant matrix, enabling fast FFT-based inversion \cite{yin10}.
Circulant matrices boast two advantages: on the mathematical side, they enable efficient matrix inversion in ADMM as discussed above; on the applications side, they are naturally involved in convolution problems, which occur, e.g., in radar imaging, Fourier optics \cite{rom09}, hyperspectral imaging \cite{mar15}, and image deblurring \cite{4776452}, as discussed below.



We experimentally show two major advantages over reference algorithms that are agnostic of circulant matrices properties.
First and foremost, our algorithms make possible to recover very large signals with just a few megabytes of GPU memory.
Second, efficient GPU memory usage improves the memory caching mechanisms effectiveness, that is the key to a speedup in excess of a tenfold factor in many scenarios.

Finally, we demonstrate our algorithms in a practical application to astronomic imaging.
We consider the reconstruction of night sky images, whose sparsity is much smaller than the number of pixels.
Due to atmospheric turbulence and optical systems imperfections (e.g, poor focusing), acquired images may be affected by small amounts of blur.
Nonlinear deblurring via inversion of a lowpass filtering is not recommended in this scenario, as it may further amplify noise.
It is however known that blurring can be incorporated in a CS framework \cite{4776452} as a convolutional filtering, so deblurring corresponds to deconvolution with a circulant matrix \cite{5495801}.
We leverage such finding and circulant sensing matrices properties to cast the problem so that the overall projections matrix is circulant as well.
Our experiments show that the implemented algorithms enable recovering a mega pixel image with a MSE in the order of $10^\textit{-2}$ in reasonable time over a desktop NVIDIA GPU.
As a final remark, our algorithms are implemented in OpenCL language albeit we experiment with NVIDIA GPUs where CUDA libraries are known to deliver better performance.
However, as a key advantage our OpenCL implementation enables deploying our algorithms over different types of GPUs (e.g., ATI) and parallel processors in general (e.g., multicore CPUs, DSPs) with little effort, as we experimentally demonstrate.

The rest of this paper is organized as follows.
Sec. \ref{sec:background} presents the relevant theoretical background on sparse signals recovery, discussing popular iterative LASSO-solving algorithms and their computational complexity.
Sec. \ref{sec:gpu} describes the architecture of a GPU, discussing the issue with memory access in matrix-vector multiplications.
In Sec. \ref{sec:cs_circulant} we introduce circulant matrices and their properties, describing an ADMM formulation that leverages such properties and is suitable for GPU parallelization.
In Sec. \ref{sec:algorithms} we present CPADMM and CPISTA, two GPU-based LASSO-solving algorithms leveraging the structure of circulant sensing matrices.
In Sec. \ref{sec:experiments} we experiment with our algorithms in recovering different types of sparse signals while in Sec.~\ref{sec:deblurring} we tackle a practical problem in compressive astronomy, namely deblurring an image in the compressed domain.
Finally, in Sec. \ref{sec:conclusions} we draw the conclusions and outline possible further developments of this work.

\section{Background}
\label{sec:background}

In this section, we provide a mathematical description of the recovery problem and of ISTA and ADMM algorithms, discussing their computational complexity.

\subsection{Sparse Signal Recovery via LASSO}
The linear system in \eqref{model} is underdetermined and has infinitely many solutions. 
As mentioned in Sec. \ref{sec:introduction}, CS theory states that imposing the sparsity constraint and with additional assumptions on the matrix $\mathbf{A}$, the following problem is well posed
\begin{equation}\label{eq:l_0}
\min\|\bm{x}\|_0 \quad\text{s.t. } \mathbf{A}\bm{x}=\bm{y}.
\end{equation}  
In fact, it can be shown \cite{can06} that, if $\bm{x}^{\star}$ is $k$-sparse and for every index set $\Gamma\subseteq\{1,\ldots,n\}$ with $|\Gamma|=2k$ the columns of $\mathbf{A}$ associated with $\Gamma$ are linearly independent, then $\bm{x}^{\star}$ is the unique solution to \eqref{eq:l_0}.  
However, this non-convex program exhibits combinatorial complexity in the size of the problem.

A popular option to solve problem \eqref{eq:l_0} is provided by convex relaxation that prescribes to minimize the following cost function, also known under the name of Least-Absolute Shrinkage and Selection Operator (LASSO, \cite{tib94}): 

\begin{equation}\label{Lasso}
\min_{\bm{x}\in\R^n}\|\bm{y}-\mathbf{A}\bm{x}\|_2^2+{2\alpha}\|\bm{x}\|_1,
\end{equation}
\noindent
where $\|\bm{y}-\mathbf{A}\bm{x}\|^2_2$ is a loss function indicating how much the vector $\bm{x}\in\R^n$ is consistent with the data $\bm{y}\in\R^m$, $\alpha>0$, and $\|\bm{x}\|_1$ is the term that promotes sparsity in the estimation.
The solution to problem \eqref{Lasso} provides an approximation of the signal with a bounded error, which is controlled by the $\alpha$ parameter (see \cite{can08, don06}).
Problem \eqref{Lasso} can be solved via a number of methods (e.g.: interior point methods, iterative methods, etc.).
In this work, we focus on iterative methods in reason of their bounded computational complexity and suitability for parallelization.
Below, we review the two well known classes of iterative LASSO-solving methods, discussing the relative pros and cons.

\subsection{Iterative Soft Thresholding Algorithm (ISTA) for LASSO}

The Iterative Soft Thresholding Algorithm (ISTA) solves the LASSO  problem \eqref{Lasso} moving at each iteration towards the steepest descent direction and applying thresholding to promote sparsity \cite{dau04}, as described in Alg.~\ref{algo:ISTA}.
In the rest of this work, we denote as $\bm{x}(t)$ the recovered signal as estimated at the end of the $t$-th iteration, $t \ge 1$.

 \begin{algorithm}[h!]
 \caption{Iterative Soft Thresholding (ISTA)}
\label{algo:ISTA}
 \begin{algorithmic}[1]
\REQUIRE{Measurements $\bm{y}\in\R^n$, sensing matrix $\mathbf{A}\in\R^{m\times n}$}
 \STATE Initialization: $\tau\in(0,\|\mathbf{A}\|_2^{-2})$, initial guess $\bm{x}(0)=0$
 \FOR{$t=1,\dots, StopIter$} 
 \STATE Residual vector computation: $$\bm{r}(t)=(\bm{y} - \mathbf{A}\bm{x}(t-1))$$
 \STATE Gradient vector computation: $$\Delta(t)= \tau \mathbf{A}^\mathsf{T}~\bm{r}(t)$$
 \STATE Soft thresholding: $$\bm{x}(t)=\eta_{\alpha}[\bm{x}(t-1)+\Delta(t)]$$
 \ENDFOR
 \end{algorithmic}
 \end{algorithm}
\noindent
Starting from an initial guess (say $\bm{x}(0)=0$), the residual vector $\bm{r}(t)$ is computed in order to evaluate how much the current estimation is consistent with the data.
The residual vector is used to compute the gradient vector: $\Delta(t)$ represents the minimizing direction of the LASSO and $\tau$ is the step-size in the update.
Finally, the $\eta_{\gamma}$ operator is a thresholding function to be applied elementwise, {\em i.e. }
\begin{equation}\label{etaS}
\eta_{\gamma}[\bm{x}]=\begin{cases}\text{sgn}(\bm{x})(|\bm{x}|-\gamma)&\text{if }|\bm{x}|>\gamma\\
0&\text{otherwise.}\end{cases}
\end{equation}
In \cite{mal10} extensive computational experiments have been conducted to optimally select parameters $\alpha$  and $\tau$. The optimization is defined in terms of phase transitions, where the number of nonzeros at which the algorithm is successful with high probability is maximized. 
If $\tau<2\|\mathbf{A}\|_2^{-2}$, then, for any initial choice $\bm{x}{(0)}$, ISTA produces a sequence $\{\bm{x}(t)\}_{t\in\N}$ which converges to a minimizer $\widehat{\bm{x}}$ of \eqref{Lasso}.

The main ISTA complexity source is in the two matrix-vector multiplications $\mathbf{A}~\bm{x}$ and $\mathbf{A}^\mathsf{T}~\bm{r}$ at lines 3 and 4 of Alg.~\ref{algo:ISTA} respectively.
Matrix-vector multiplications are well-suited for GPU parallelization due to the intrinsic independence of each row-vector product.
However, ISTA may require a large number of iterations to converge to recover the signal.
Moreover, accessing large matrices in GPU memory represents a bottleneck to the actual GPU parallelism, as we discuss in the following.

\subsection{Alternating Direction Method of Multipliers (ADMM) for LASSO}

The Alternating Direction Method of Multipliers (ADMM, \cite{boy10}) is another well-known method for solving~\eqref{Lasso}. 
The optimization problem is reformulated as follows:

\begin{equation}\label{Lasso2}
\min_{\bm{x}\in\R^n}\|\bm{y}-\mathbf{A}\bm{x}\|_2^2+{2\alpha}\|\bm{z}\|_1,\qquad \text{s.t. }\bm{x}-\bm{z}=0.
\end{equation}
\noindent
ADMM attempts to blend the benefits of dual decomposition and augmented Lagrangian methods for constrained optimization by minimizing the augmented Lagrangian in a iterative way with respect to the primal variable $\bm{x}$ and the dual variables.
More formally, ADMM addresses the optimization problem

\begin{equation}\label{d2lp_Lag}
\min_{\bm{x}\in\R^n}\|\bm{y}-\mathbf{A}\bm{x}\|_2^2+{2\alpha}\|\bm{z}\|_1+\bm{u}^{\mathsf{T}}(\bm{x}-\bm{z})+{\rho}\|\bm{x}-\bm{z}\|^2
\end{equation}
\noindent
where $\bm{u}$ is the dual variable or Lagrange multiplier related to the constraint in \eqref{Lasso2} and $\bm{z}$ is an auxiliary vector.
The ADMM algorithm is summarized in pseudo-code as Alg.~\ref{algo:ADMM}.

 \begin{algorithm}[h!]
 \caption{Alternating Direction Method of Multipliers (ADMM)}
 \label{algo:ADMM}
 \begin{algorithmic}[1]
\REQUIRE{Measurements $\bm{y}\in\R^n$, sensing matrix $\mathbf{A}\in\R^{m\times n}$}
 \STATE Initial guess $\bm{x}(0)=\bm{u}(0)=\bm{z}(0)=0$
 \STATE Initial inversion: $$\mathbf{B} = (\mathbf{A}^\mathsf{T}\mathbf{A} + \rho \mathbf{I})^{-1}$$
 \FOR{$t=1,\dots, StopIter$}
 \STATE Primal variables update: $$\bm{x}(t)= \mathbf{B} (\mathbf{A}^\mathsf{T}\bm{y} + \rho(\bm{z}(t-1) - \bm{u}(t-1)))$$
 \STATE Soft thresholding step: $$\bm{z}(t)=\eta_{\alpha/\rho}[\bm{x}(t) + \bm{u}(t-1)]$$
 \STATE Dual variables update: $$\bm{u}(t)=\bm{u}(t-1)+\bm{x}(t)-\bm{z}(t)$$
 \ENDFOR
 \end{algorithmic}
 \end{algorithm}
\noindent
ADMM consists of an initial stage where matrix $\mathbf{A}^\mathsf{T}\mathbf{A} + \rho \mathbf{I}$ is inverted (line 2) and of an iterative stage (rest of the algorithm).
The convergence of ADMM can be established in several ways \cite{boy10}.

Complexity-wise, ADMM converges in fewer iterations than ISTA, while each iteration complexity is comparable.
In fact, each ADMM iteration requires performing two matrix-vector multiplications at line 4 of \ref{algo:ADMM} and each multiplication can be parallelized.
However, ADMM initially requires inverting and storing in memory the $n\times n$ matrix $\mathbf{A}^\mathsf{T}\mathbf{A}$.
The inversion complexity is $O(n^3)$, so it becomes the dominant complexity source as the sampled signal dimension $n$ increases.
In the following, we show how circulant matrices can be leveraged to keep such complexity under control.

\section{GPU Architecture}
\label{sec:gpu}

This section first describes a GPU hardware architecture and the OpenCL parallel programming model.
We focus in particular on matrix-vector multiplications, which are the main complexity source in the previously described LASSO-solving algorithms as discussed in the following section.

\subsection{Hardware Architecture}

\begin{figure}[h]
\begin{minipage}[b]{1.0\linewidth}
\vspace{-0.25 cm}
  \centering
  \rotatebox{0} {
	\includegraphics[width=0.8\columnwidth]{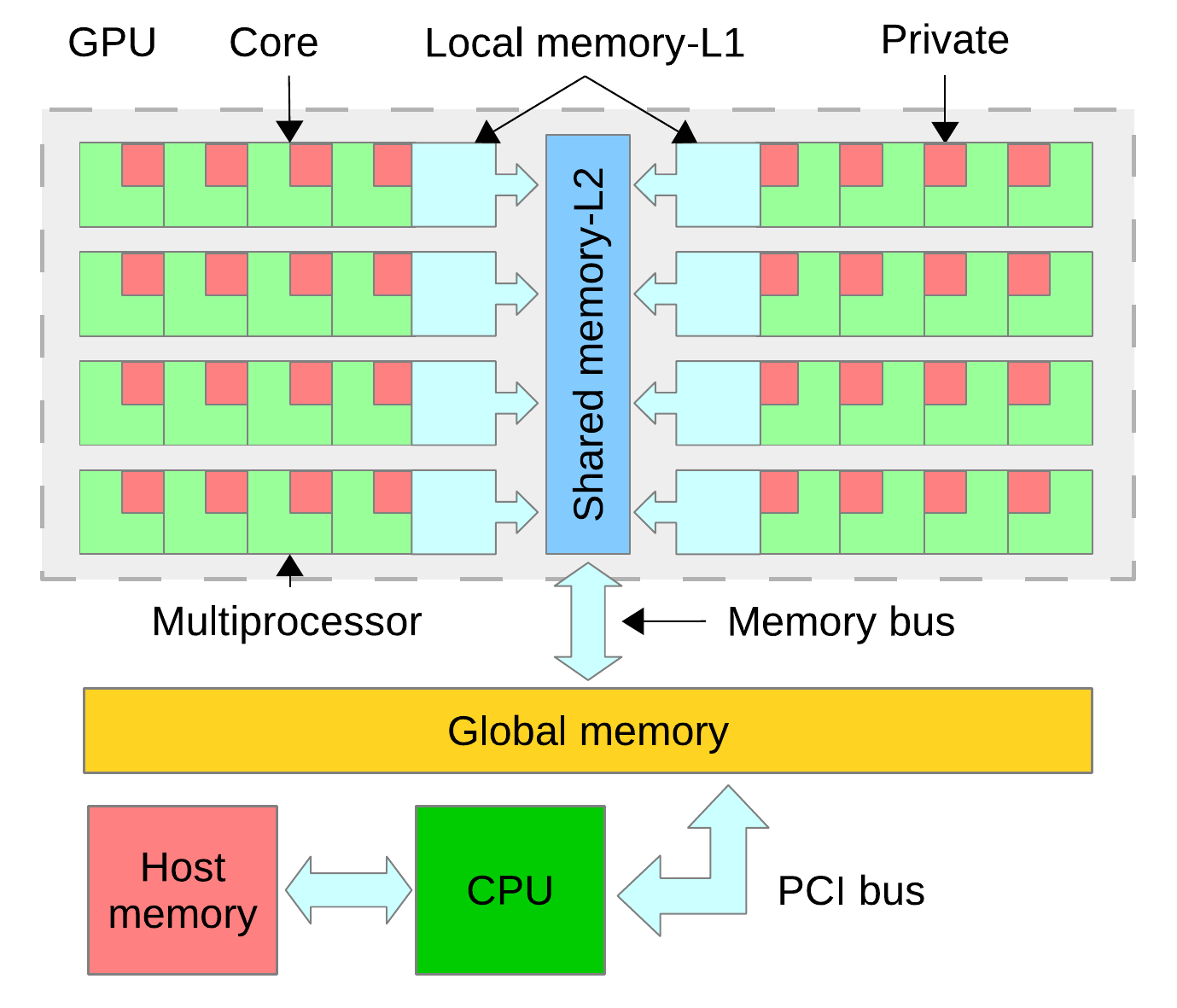}
  }
\end{minipage}
\caption{Simplified GPU architecture with relative memory. Notice that the memory bus between the GPU and the global memory is shared by all the GPU cores and represents the main performance bottleneck in accessing the global memory.}
\label{fig:gpu_architecture}
\end{figure}

Figure~\ref{fig:gpu_architecture} shows that a modern GPU includes two main components.
The first component is the parallel processor, simply referred to as GPU in the following (dashed box in the figure).
The GPU basic computing unit is the \emph{core} and each core includes a small amount of \emph{private} memory accessible exclusively to it (e.g., to hold temporary variables).
Groups of cores are physically arranged into \emph{multiprocessor} (MPs), and each MP includes some tens of kilobytes \emph{local} memory accessible by all the MP cores.
All MPs access a large (up to several gigabytes) off-chip \emph{global} memory through a shared \emph{memory bus}, which is the second main component of a GPU.
GPUs typically include a second-level (L2) cache memory shared by all MPs to hide the global memory access latency (e.g., 2MB in Maxwell-based GPUs).
Latency is defined here as the average time (or number of processor or bus cycles) required by a core to access an operand in the global memory.
Global memory is the only memory area accessible both to the GPU and to the main CPU, also known as \emph{host} processor.
The host processor accesses the global memory via the PCI bus and can move data between the main memory, known as \emph{host} memory, and the global memory. 

\subsection{OpenCL Programming Language}

OpenCL~\cite{opencl} is a C-like language for programming parallel computers such as GPUs.
The typical OpenCL programming model follows the pattern below.
\\
Preliminarily, one or more OpenCL \emph{kernels} and the relative input data (e.g., the sensing matrix and the sampled signal vector) are copied to the GPU global memory.
OpenCL kernels are primitive, concurrent, functions taking care of one elementary task each: for example multiplying one row of the sensing matrix by the sampled signal vector.
\\
Next, the host processor issues a number of independent kernel instances called \emph{work-items} for execution on the GPU, and waits for all the instances to yield, i.e. to complete.
Each work item is a logical entity that is uniquely identified GPU-wide through its \emph{global identifier}.
A scheduler resident on the GPU distributes the work-items among the physical computing cores.
In the case of a kernel performing one matrix-vector multiplications, a work-item may be issued for each row-vector multiplication.
Each work-item fetches the (relevant part of the) input data from the global memory, performs the relevant computation and stores the output into the global memory.
After all work-items have terminated, the host processor moves the output from the global memory back to the host memory, completing the execution of the OpenCL program.

\section{Algorithmic Design with Circulant Sensing Matrices}
\label{sec:cs_circulant}

In this section we discuss the issues with accessing GPUs memory, motivating the use of circulant matrices for solving the LASSO problem and proposing an ADMM formulation that relies on circulant matrices.

\subsection{The Issue with GPU Memory}

A key issue in designing high performing GPU-accelerated algorithms is making efficient use of the GPU memory architecture described in the previous section.
Other than fitting the algorithm working set in the available GPU memory, the goal is to minimize the global memory access latency, which is a major performance bottleneck.
Memory access latency is especially a problem in matrix-vector multiplications, which represent the main complexity source in LASSO-solving algorithms.
Consider for example the multiplication $\bm{y} = \mathbf{A} \bm{x}$ where the sensing matrix $\mathbf{A}$ is stored in global memory in standard row-major order.
Let us assume that an OpenCL kernel performs the multiplication row-wise, i.e. each $i$-th work-item multiplies the $i$-th row of $\mathbf{A}$ by $\bm{x}$.
Each $i$-th work-item accesses a different row of $\mathbf{A}$, i.e. a distinct consecutive set of location in the global memory.
Therefore, the GPU cores will compete for accessing the global memory over the shared bus, stalling until the memory bus becomes free.
The fast on-chip caches may help to buffer frequently accessed memory areas, reducing latency.
However, caches are limited in size and may be unable to hold entirely large data structures such as the whole matrix $\mathbf{A}$.
Namely, when the signal $n$ increases, structures such as the sensing matrix are increasingly unlikely to fit inside the GPU caches, memory access latency increases and the GPU performance drops.
A number of good practices such as reducing the number of transfers between host and global memory and performing coalesced memory accesses help to this end.
However, such practices cannot set completely off the memory access bottleneck in matrix-vector multiplications.
In the following, we workaround such issue leveraging circulant matrices and their properties.

\subsection{Circulant Sensing Matrices}

Given a generic dense matrix $\mathbf{A}$ of size $m\times n$ and a vector $\bm{x}$ of size $n$, the product requires $O(mn)$ computations.
Moreover the multiplication has to be performed over and over again with different input vectors.
In order to reduce the storage and computational complexity, one can consider structured matrices which are dense but depend on only $O(n)$ parameters. 
In particular, circulant random matrices are almost as effective as the Gaussian random matrix for the purpose of recovering compressed signals~\cite{ra10}.

Let $\bm{v}$ be the $1 \times n$ vector corresponding to the first $\mathbf{A}$ row.
In the following, we refer to vector $\bm{v}$ as \emph{sensing vector}: $\mathbf{A}$ is circulant, so each $\mathbf{A}$ row can be expressed as a shift of the sensing vector.
Namely, $\mathbf{A}_{i,j} = \bm{v}_{(j-i) \bmod(n)}$, where $\bmod$ defines the remainder operator in the rest of this manuscript.
Similarly, each column of the transpose $\mathbf{A}^\mathsf{T}$ can be expressed as a shift of the sensing vector.
Namely, $\mathbf{A}^\mathsf{T}_{i,j} = \mathbf{A}_{j,i}$, so $\mathbf{A}^\mathsf{T}_{i,j} = \bm{v}_{(j-i) \bmod(n)}$. 

From a computational viewpoint, circulant matrices can be diagonalized using the discrete Fourier transform \cite{yin10}.
So, the matrix-vector multiplication $y = \mathbf{A}~x$ can be efficiently performed using the Fast Fourier Transform (FFT) with a complexity of order $O(n \log(n))$.
While the use of the circulant matrices reduces naturally the storage for ISTA, ADMM need still to invert and store an $n\times n$ matrix $\mathbf{A}^\mathsf{T}\mathbf{A}$.
In the following we review the basic ideas for an ADMM with partial circulant matrices.

\subsection{ADMM for Circulant Matrices}

In \cite{yin10} a fast ADMM is proposed for partial circulant matrices.
From now on we consider the Lasso problem in \eqref{Lasso} with random partial circulant sensing matrix.
Let $\Omega\subseteq\{1,\ldots, n\}$, chosen at random, with $|\Omega|=m$, then we consider $\mathbf{A}$ of the form $\mathbf{A}=\mathbf{P}\mathbf{C}$ where $\mathbf{C}\in\R^{n\times n}$ is circulant square and $\mathbf{P}\in\R^{m\times n}$ is a binary diagonal matrix, with $P_{i,i}=1$ if $i\in\Omega$.
\\
The LASSO problem can be  written as follows:
\begin{align*}
    \frac{1}{2}\left\|\bm{y}-\mathbf{P} \bm{v} \right\|_2^2 +\alpha \|\bm{z}\|_1\\
\text{s.t. }\bm{v}=\mathbf{C}\bm{x},\ \bm{z}=\bm{x}.
\end{align*}
\noindent
Consider now the augmented Lagrangian function

\begin{align*}
L(\bm{x},\bm{v},\bm{z},\bm{\mu}, \bm{\bm{\nu}})&=\frac{1}{2}\left\|\bm{y}-\mathbf{P} \bm{v} \right\|_2^2 +\alpha \|\bm{z}\|_1 +\sigma \bm{\nu}^{\mathsf{T}}(\bm{x}-\bm{z})\\
&+ \frac{\sigma}{2}\left\|\bm{x}-\bm{z}\right\|_2^2+\rho\bm{\mu}^{\mathsf{T}} (\bm{v}-\mathbf{C}\bm{x})+\frac{\rho}{2}\left\|\bm{v}-\mathbf{C}\bm{x} \right\|_2^2
\end{align*}
\noindent
where $\bm{x}$, $\bm{v}$, $\bm{z}$ are primal variables and $\bm{\mu}$, $\bm{\nu}$ are the Lagrangian multipliers.
By minimizing $L(\bm{x},\bm{v},\bm{z},\bm{\mu}, \bm{\nu})$ in an iterative fashion, we obtain the following updates: start from an initial condition $\bm{\mu}(0)=\bm{\nu}(0)=\bm{z}(0)=\bm{v}(0)=0$

\begin{gather*}
(\bm{x}(t+1),\bm{v}(t+1),\bm{z}(t+1))=\argmin_{x}L(\bm{x},\bm{v},\bm{z},\bm{\mu}(t), \bm{\nu}(t))\\
\bm{\mu}(t+1)=\bm{\mu}(t)+\tau_1\nabla_{\bm{\mu}}L(\bm{x}(t+1),\bm{v}(t+1),\bm{z}(t+1),\bm{\mu}, \bm{\nu})\\
\bm{\nu}(t+1)=\bm{\nu}(t)+\tau_2\nabla_{\bm{\nu}}L(\bm{x}(t+1),\bm{v}(t+1),\bm{z}(t+1),\bm{\mu}, \bm{\nu})
\end{gather*}
\noindent
that lead to pseudocode in Algorithm \ref{algo:CPADMM}.

 \begin{algorithm}[h!]
 \caption{Circulant ADMM}
 \label{algo:CPADMM}
 \begin{algorithmic}[1]
\REQUIRE{Measurements $y\in\R^n$, sensing matrix $\mathbf{A}\in\R^{m\times n}$, $\mathbf{A}=\mathbf{P}\mathbf{C}$, $\mathbf{P}\in\{0,1\}^{m\times n}$, $\mathbf{C}\in\R^{n\times n}$ circulant}
 \STATE Initialization: $\bm{\mu}(0)=\bm{\nu}(0)=\bm{z}(0)=\bm{v}(0)=0$
 \STATE Initial inversion: $$\mathbf{B} = (\rho \mathbf{C}^{\mathsf{T}} \mathbf{C} +\sigma \mathbf{I})^{-1},~~ \mathbf{D}=( \mathbf{P}^{\mathsf{T}} \mathbf{P} +\rho \mathbf{I})^{-1} $$
 \FOR{$t=1,\dots, StopIter$} 
 \STATE Primal variables update: 
 \begin{align*}
 \bm{x}(t)&=\mathbf{B} (\rho \mathbf{C}^{\mathsf{T}} \bm{v}(t-1)+\sigma(\bm{z}(t-1)-\bm{\nu}(t-1)))\\
 \bm{v}(t)&=\mathbf{D}(\rho \mathbf{C} \bm{x}(t)-\rho \bm{\mu}(t-1)+\mathbf{P}^{\mathsf{T}}y)
 \end{align*}
 \STATE Soft-Thresholding step:
 $$\bm{z}(t)=\eta_{\alpha/\sigma}[\bm{x}(t)+\bm{\nu}(t-1)]$$
 \STATE Dual variables update:
 \begin{align*}
 \bm{\mu}(t)&=\bm{\mu}(t-1)+\tau_1(\bm{v}(t)-\mathbf{C}\bm{x}(t))\\
 \bm{\nu}(t)&=\bm{\nu}(t-1)+\tau_2(\bm{x}(t)-\bm{z}(t))\\
 \bm{v}(t) &= \bm{v}(t)+\bm{\mu}(t)
\end{align*}
 \ENDFOR
 \end{algorithmic}
 \end{algorithm}
\noindent
The algorithm is shown to converge for $\tau_1=\tau_2\in(0,(\sqrt{5}+1)/2)$ in \cite{fou13}.

We recall that the inversion of both matrices $(\rho \mathbf{C}^{\mathsf{T}} \mathbf{C} +\sigma \mathbf{I})$ and $ \mathbf{P}^{\mathsf{T}} \mathbf{P} +\rho \mathbf{I}$  can be performed offline.
However, we point out that $(\rho \mathbf{C}^{\mathsf{T}} \mathbf{C} +\sigma \mathbf{I})\in\R^{n\times n}$ is a square circulant matrix, so it can be inverted via FFT, which reduces the computational cost from $O(n^3)$ to $O(n\log n)$. This works around the main complexity element of ADMM as described in Alg. \ref{algo:CPADMM}.
$\mathbf{P}^{\mathsf{T}} \mathbf{P} -\rho \mathbf{I}$ is a diagonal matrix with $i$-th element in the diagonal equal to $1+\rho$ if $i\in\Omega$ and $\rho$ otherwise.
Moreover and most important, the inverted matrices $\mathbf{B}$ and $\mathbf{D}$ are still circulant, thus they benefit from the properties of circulant matrices as described above.

We leverage such circulant matrices properties to design efficient, GPU-accelerated, matrix-vector multiplication algorithms that we discuss in detail in the next section.

\section{GPU Algorithms}
\label{sec:algorithms}

\begin{figure}
  \begin{center}
  \rotatebox{0}{
  \includegraphics[width=0.95\columnwidth]{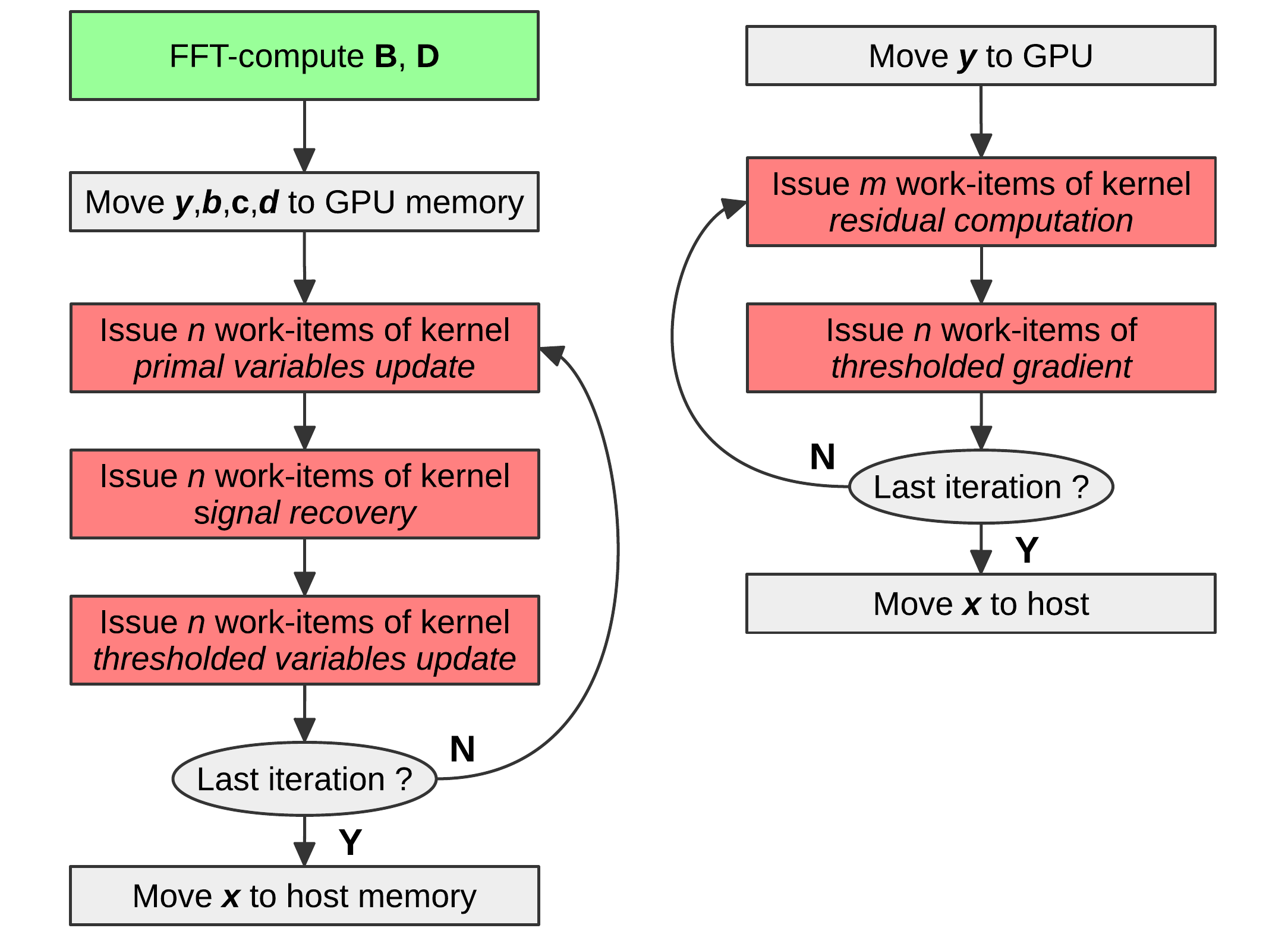}
  }
  \end{center}
  \caption{CPADMM (left) and CPISTA (right): green boxes correspond to code executed by the CPU, red boxes to code execute by the GPU as OpenCL kernels.}
  \label{alg:cpista_outline}
\end{figure}

In this section we describe CPADMM, a circulant parallel ADMM algorithm, and  CPISTA, a circulant, parallel ISTA algorithm.
The algorithms rely on the GPU for the most computationally-intensive tasks as illustrated in Fig.~\ref{alg:cpista_outline}.
Both algorithms rely on multiple distinct kernels to cope with the hardware limitations of some GPUs.
Namely, multiple kernels are required to allow multiple MPs to yield and synchronize properly.
Each kernel executes only one matrix-vector multiplication to guarantee the correct kernel execution order.
All kernels below are described in pseudo-OpenCL language.

\subsection{CPADMM}
\label{sec:cpadmm}

This section describes CPADMM (Circulant Parallel ADMM), a GPU-optimized version of the Circulant ADMM algorithm we presented as Alg.~\ref{algo:CPADMM}, as outlined as a flow-chart in Fig.~\ref{alg:cpista_outline} (left).

First, the inverted matrices $\mathbf{B} = (\rho \mathbf{C}^{\mathsf{T}} \mathbf{C} +\sigma \mathbf{I})^{-1}$ and $\mathbf{D} = ( \mathbf{P}^{\mathsf{T}} \mathbf{P} +\rho \mathbf{I})^{-1}$ are computed over the CPU (line 2 of Alg.~\ref{algo:CPADMM}).
Both $\rho \mathbf{C}^{\mathsf{T}} \mathbf{C} +\sigma \mathbf{I}$ and $\mathbf{P}^{\mathsf{T}} \mathbf{P} +\rho \mathbf{I}$  are $n \times n$ circulant matrices.
It is well known that square circulant matrices are diagonalizable by the Fourier matrix, which reduces their inversion to the inversion of the associated diagonal eigenvalue matrix and two FFTs \cite{yin10}.
Such preliminary inversions are not performed on the GPU because their FFT-complexity grows only linearly with $n$ and thus can be efficiently performed over the CPU, as we experimentally show later on.
Then, $\mathbf{B}$ and $\mathbf{D}$ are $n \times n$ matrices and their footprint would grow with the square of $n$.
However, $\mathbf{B}$ is circulant and is completely represented by its first row that in the following we denote as the $n$-elements vector $\bm{b}$.
Similarly, $\mathbf{D}$ is diagonal, so is represented by its diagonal that we indicate as the $n$-elements vector $\bm{d}$.
Therefore, at the end of line 2 in Alg.~\ref{algo:CPADMM} we represent $\mathbf{B}$ and $\mathbf{D}$ in the CPU memory as two $n$-elements vectors $\bm{b}$ and $\bm{d}$.
Notice that $\mathbf{C}$ and $\mathbf{C}^{\mathsf{T}}$ are circulant as well, and both are represented by $\mathbf{C}$ first row that we indicate as as the $n$-elements vector $\bm{c}$.
\\
Next, the required input data is prepared in the GPU memory (in the following, we refer to the GPU \emph{global} memory simply as GPU memory).
First, the four $n$-elements vectors $\bm{y}$, $\bm{b}$, $\bm{c}$ and $\bm{d}$ are copied from the CPU memory to the GPU memory.
Then, the $n$-elements vectors $\bm{\mu}$, $\bm{\nu}$, $\bm{z}$ and $\bm{v}$ are allocated in the GPU memory and initialized to zero.
Finally, we allocate in the GPU memory one $n$-elements vector $\bm{\beta}$ to hold temporary results between successive kernel calls.
Once all the required data structures are allocated in the GPU memory, the first CPADMM algorithm iteration is executed on the GPU.
At each $t$-th iteration, three distinct OpenCL kernels are issued in sequence.
The core of each kernel is one of the three matrix-vector multiplications required at each ADMM iteration as detailed in the following.

The Kernel in Alg.~\ref{alg:cpadmm_1} takes care of the first part of the primal variables update stage at line 4 of Alg.~\ref{algo:CPADMM} and $n$ instances thereof are issued.
Throughout the rest of this paper, we indicate a work-item \emph{global identifier} (i.e., the unique identifier of a kernel instance as discussed in Sec.~\ref{sec:gpu}) with the letter $i$.
Namely, at the $t$-th iteration of the algorithm, the $i$-th work-item multiplies the $i$-th row of matrix $\mathbf{C}$ (indicated as $\mathbf{C}_i$ below) by the vector $\bm{v}$ computed at the previous iteration.
Matrix $\mathbf{C}$ is actually represented by $\bm{c}$ in GPU memory, so each work-item multiplies the $(i-j)\bmod(n)$-th element of $\bm{c}$ by the $j$-th element of $\bm{v}$ during the $for$ loop storing the partial results in the private accumulator $s$.
Then, the kernel multiplies the accumulator by $\rho$, adds the $i$-th element of $\bm{z}$, subtracts the $i$-th element of $\bm{v}$ and stores the result back in the GPU memory as the $i$-th elements of the auxiliary vector $\bm{\beta}$.
Being $\bm{\beta}$ allocated in GPU memory, it that can be accessed later on by the other kernels after all $m$ work-items of the kernel have yielded.

\begin{algorithm}[h]
  \caption{CPADMM: \emph{Primal Variables Update} kernel}
  \label{alg:cpadmm_1}
  \begin{algorithmic}
  \STATE $i \gets {get\_global\_id}(0)$
  \STATE $s \gets 0$
  \FOR{$j = 1 \to n$}
    \STATE $s \gets s + c_{(i - j) \bmod (n)} ~ v_j(t-1)$
  \ENDFOR
  \STATE $\beta_i(t) \gets s~\rho + z_i(t-1) - \nu_i(t-1)$
  \end{algorithmic}
\end{algorithm}

After all work-items of the kernel in Alg.~\ref{alg:cpadmm_1} have yielded, $n$ work-items of the kernel in Alg.~\ref{alg:cpadmm_2} are issued.
The kernel core task is performing the second multiplication required by the primal variables update stage at line 4 of Alg.~\ref{algo:CPADMM} and estimating the original signal $x$.
Each work-item multiplies the $i$-th row of matrix $\mathbf{B}$ by the temporary vector $\bm{\beta}$ computed by kernel~\ref{alg:cpadmm_1}.
We recall that $\mathbf{B}$ is circulant and is represented by the $n$-elements vector $\bm{b}$.
Each work-item performs the vectorial multiplication between each $j$-th element of vector $\bm{b}$ and vector $\bm{\beta}$ over a $for$ loop.
Each of the $n$ work-items finally updates the $i$-th element of the estimated signal to recover $\bm{x}$ in global memory.

\begin{algorithm}[h]
  \caption{CPADMM: \emph{Signal Recovery} kernel}
  \label{alg:cpadmm_2}
  \begin{algorithmic}
  \STATE $i \gets {get\_global\_id}(0)$
  \STATE $s \gets 0$
  \FOR{$j = 1 \to n$}
    \STATE $s \gets s + b_{(j - i) \bmod (n)} ~ \beta_j(t)$
  \ENDFOR
  \STATE $x_i(t) \gets s$
  \end{algorithmic}
\end{algorithm}

Finally, $n$ work-items of the kernel in Alg.~\ref{alg:cpadmm_3} are issued, completing one CPADMM iteration.
\\
First, each work-item multiplies the $i$-th row of matrix $\mathbf{C}$ by the recovered signal $\bm{x}$ over a $for$ loop.
As $\mathbf{C}$ is circulant, the kernel actually loops over vector $\bm{c}$ elements and stores the partial results in the private accumulator $s$.
Then, the $i$-th element of the dual variable $\bm{\mu}$ is subtracted from $s$ and is multiplied by $\rho$, updating $s$.
Next, the kernel multiplies the $(i - j) \bmod (n)$-th element of $\bm{p}$ by the $i$-th element of $\bm{y}$ and updates $s$.
Finally, the kernel multiplies $s$ by the $(i - j) \bmod (n)$-th element of $\bm{d}$, updating the $i$-th element of the primary variable $\bm{v}$ in GPU memory.
\\
Second, the kernel performs soft thresholding ($\eta$ operator) over $x_i + \nu_i$ and updates the $i$-th element of $\bm{z}$ in the GPU memory.
\\
Third and last, the kernel computes and updates the $i$-th element of vectors $\bm{\mu}$, $\bm{\nu}$ and $\bm{v}$ in GPU memory, concluding one CPADMM iteration.

\begin{algorithm}[h]
  \caption{CPADMM: \emph{Thresholded Variables Update} kernel}
  \label{alg:cpadmm_3}
  \begin{algorithmic}
  \STATE $i \gets {get\_global\_id}(0)$
  \STATE $s \gets 0$
  \FOR{$j = 1 \to n$}
    \STATE $s \gets s + c_{(j - i) \bmod (n)} ~ x_j(t)$
  \ENDFOR
  \STATE $s \gets \rho (s - \mu_i(t-1))$
  \STATE $s \gets s + (p_{(i - j) \bmod (n)}~ y_i)$
  \STATE $v_i(t) \gets s~d_{(i - j) \bmod (n)}$
  \STATE $z_i(t) \gets \eta_{\alpha/\sigma}~[x_i(t) + \nu_i(t-1)]$
  \STATE $\mu_i(t) \gets v_i(t) - s$
  \STATE $\nu_i(t) \gets x_i(t) - z_i(t)$
  \STATE $v_i(t) \gets v_i(t-1) - \mu_i(t)$
  \end{algorithmic}
\end{algorithm}

\subsection{CPISTA}
\label{sec:cpista}

Then we describe CPISTA (Circulant Parallel ISTA), a GPU-accelerated version of the ISTA algorithm ~\ref{algo:ISTA} that leverages circulant sensing matrices as outlined in the flow-chart in Fig.~\ref{alg:cpista_outline} (right).
Refraining the notation of Sec.~\ref{sec:cs_circulant}, we indicate the sensing matrix $A$ first row as the $n$-elements vector $\bm{v}$.
First, $\bm{v}$ and the $m$-elements vector $\bm{y}$ (i.e., the sampled signal) are copied from the host memory to the GPU memory.
Then, the $m$-elements vector $\bm{r}$ is allocated in the GPU memory to hold the residuals.
Also, an $n$-elements vector $\bm{x}$ is allocated in global memory and initialized to zero to hold the recovered signal.
Next, the first CPISTA iteration takes place.


The kernel in Alg.~\ref{alg:cpista_1} takes care of computing the \emph{residuals vector} $\bm{r}(t)$.
The kernel takes in input vectors $\bm{y}$, $\bm{v}$ and $\bm{x}$, the latter holding the signal $\bm{x}(t-1)$ recovered during the previous iteration ($\bm{x}(t-1) = 0$ for $t=1$).
With reference to Alg.~\ref{algo:ISTA}, line 3, the each $i$-th work-item computes the $i$-th element of $\bm{r}(t)$.
Each work-item multiplies the $i$-th row of $\mathbf{A}$ by $\bm{x}(t-1)$ and subtracts the $i$-th $\bm{y}$ component from it).
So, a total of $m$ work-items are issued on the GPU to compute the residuals vector $\bm{r}(t)$.
Each work-item multiplies the $j$-th element of $\bm{x}$ by the $(i+j) \bmod (n)$-th element of $\bm{v}$.
Accumulator $s$ holds the partial results of the inner product computed during the $j$-indexed $for$ loop over the $n$-elements of $\mathbf{A}_i$ and $\bm{x}$.
The residual vector $i$-th element $r_i$ is finally stored in global memory, so that can be accessed later on by Kernel~\ref{alg:cpista_2} after all $m$ work-items have yielded.

\begin{algorithm}
  \caption{CPISTA: \emph{Residual Computation} kernel}
  \label{alg:cpista_1}
  \begin{algorithmic}
  \STATE $i \gets {get\_global\_id}(0)$
  \STATE $s \gets 0$
  \FOR{$j = 1 \to n$} 
    \STATE $s \gets s + v_{i+j \bmod (n)} ~ x_j(t-1)$
  \ENDFOR
  \STATE $r_i(t) \gets y_i - s$
  \end{algorithmic}
\end{algorithm}

The kernel in Alg.~\ref{alg:cpista_2} performs a soft thresholding and takes care of updating the gradient.
The kernel takes in input vectors $\bm{v}$, $\bm{x}$ and $\bm{r}$, the latter holding the residuals $\bm{r}(t)$ computed by Alg.~\ref{alg:cpista_1}.
As $\bm{x}$ has dimension $n$, $n$ work-items are issued in parallel.
\\
First, with reference to Alg.~\ref{algo:ISTA} at line 4, the kernel computes the $i$-th element of the gradient vector $\Delta_i(t)$.
The residual element $\Delta_i(t)$ is computed as the a vectorial multiplication between $\mathbf{A}_i^\mathsf{T}$ and the residual vector $\bm{r}(t)$.
$\mathbf{A}^\mathsf{T}$ is column-circulant, so $\mathbf{A}^\mathsf{T}_{i,j} = \mathbf{A}_{j,i} = v_{(j+i) \bmod (n)}$: i.e., the kernel accesses the sensing vector $\bm{v}$ starting from position $(j+i)$.
The multiplication is performed within the $for$ loop, accumulator $s$ holding the partial results.
Then, $s$ is multiplied by the thresholding gradient value $\tau$, obtaining the actual $\Delta_i(t)$.
\\
Second, with reference to Alg.~\ref{algo:ISTA} at line 5, the kernel updates the estimate of the $i$-th element of the recovered signal $\bm{x}(t)$.
The $i$-th element of signal $\bm{x}$ recovered at the previous iteration, $\bm{x}(t-1)$ is added to the gradient vector $i$-th element held in $s$.
Then, the soft thresholding operator in \eqref{etaS} is applied to accumulator $s$.
The output of the thresholding operator is stored in global memory as the $i$-th element of the updated recovered signal $\bm{x}(t)$ before the kernel yields.

\begin{algorithm}
  \caption{CPISTA: \emph{Thresholded Gradient} kernel}
  \label{alg:cpista_2}
  \begin{algorithmic}
  \STATE $i \gets {get\_global\_id}(0)$
  \STATE $s \gets 0$
  \FOR{$j = 1 \to m$}
    \STATE $s \gets s + v_{i+j \bmod (n)} ~ r_j (t)$
  \ENDFOR
  \STATE $s \gets s ~ \tau + x_i(t-1)$
  \STATE $\widetilde {x}_i(t) \gets \eta_{\alpha}~[s] $
  \end{algorithmic}
\end{algorithm}

With respect to the PISTA algorithm we described in~\cite{pcs2013}, CPISTA boasts the same degree of nominal parallelism but leverages the properties of circulant matrices with the advantages shown in the next section.

\subsection{PADMM}
\label{sec:padmm}

Finally, we describe PADMM, a hybrid GPU-CPU parallel embodiment of Alg.~\ref{algo:ADMM} that we use to benchmark CPADMM.
Differently from CPADMM, PADMM does not leverage circulant matrices for efficient representation in GPU memory, but otherwise boasts the same degree of nominal parallelism.
Therefore, PADMM allows to assess the impact of efficient circulant matrices representation in GPU memory for the same degree of execution parallelism.

First, the vectors corresponding to the initial solution $\bm{u}(0) = \bm{z}(0) = 0$ are initialized by the CPU and moved to the GPU memory.
Next, the $n \times n$ symmetric positive definite  matrix $\mathbf{B} = \mathbf{A}^\mathsf{T} \mathbf{A}+\rho \mathbf{I}$ (line 1 of Alg.~\ref{algo:ADMM}) is precomputed on the CPU and is stored in the GPU memory.
Then, vector $\mathbf{A}^\mathsf{T}~\bm{y}$ (line 4) is constant over time so is precomputed on the GPU and stored in GPU memory.
Similarly, $\mathbf{A}^\mathsf{T}\bm{y} + \rho (\bm{z}(t-1) - \bm{u}(t-1))$ (line 4) is precomputed on the CPU and stored in the GPU memory as the $n \times 1$ vector $\mathbf{C}(t)$.
As the initial solution is $\bm{u}(0) = \bm{z}(0) = 0$, we have that $\mathbf{C}(0) = \mathbf{A}^\mathsf{T} \bm{y}$.

PADMM implements the rest of Alg.~\ref{algo:ADMM} on the GPU via two OpenCL kernels as follows.
\\
The first kernel takes as input matrix $\mathbf{B}$, vector $\mathbf{C}(t)$ and the dual variables computed at the previous iteration $\bm{u}(t-1)$, $\bm{z}(t-1)$.
A number of $n$ work-items of the kernel are issued in parallel.
The kernel updates the $i$-th element of the recovered signal $x_i(t)$ and of the dual variables $u_i(t)$ and $z_i(t)$.
\\
The second kernel updates vector $\mathbf{C}(t)$, computing the $i$-the element of $\mathbf{C}$ as $\mathbf{C}_i =  \rho~(z_i - u_i)~~+~~(\mathbf{A}^\mathsf{T}~\bm{y})_i$.
A number of $n$ work-items of the kernel are issued in parallel.


\section{Experiments}
\label{sec:experiments}

In this section, we experiment with our proposed algorithms recovering a sparse signal $\bm{x}^{\star}$ of size $n$ where $\bm{x}^{\star}$ is randomly generated so that $k \le n$ elements are different from zero, and each of them is drawn from a Gaussian distribution.
Signal $\bm{x}^{\star}$ is sampled  as $\bm{y} = \mathbf{A}\bm{x}^{\star}$, where $\mathbf{A}$ is an $m \times n$  circulant matrix with entries drawn from a standard Gaussian distribution.
In all of the following experiments, we consider $m = \frac{n}{2}$ and $k \simeq \frac{n}{10}$.

Our experimental testbed is a four-way workstation for a total of 32 CPU cores and equipped with 32 GB of RAM.
The workstation is also equipped with an OpenCL-capable NVIDIA Titan X GPU, which includes 3072 CUDA cores based on the \emph{Maxwell} microarchitecture and 12 GB of memory.

In our experiments we consider multiple performance metrics.
First, we use as recovery error metrics the mean square error (MSE), which is defined as:

$$\text{MSE}=\frac{\left\|\bm{x}^{\star}-\bm{x} \right\|_2^2}{n}$$ where $\bm{x}$ is the estimated recovered signal.
\noindent
We consider the signal \emph{recovered} when the recovery error falls below a target error equal to $10^{-4}$.
Therefore, we define \emph{recovery time} the time required to recover the signal with an MSE equal or lower to $10^{-4}$.
Second, we define \emph{memory footprint} the amount of GPU memory required by a particular algorithm to recover our test signal.
The footprint is logged at runtime using the \textit{nvidia-smi} tool, which measures the actual amount of allocated GPU memory.
In all the following experiments, we set $\alpha=10^{-4}$ and $\sigma=\tau=10^{-1}$, which enables a reasonable tradeoff between probability to recover the signal and recovery speed.
In all the below experiments, CPADMM identifies our parallel, circulant-enabled, ADMM version we proposed in Sec.~\ref{sec:cpadmm}.
The PADMM curve corresponds instead to the parallel, circulant-unaware, ADMM reference implementation we described in Sec.~\ref{sec:padmm}.
Similarly,the CPISTA curve corresponds to our parallel, circulant-enabled, ISTA version we proposed in Sec.~\ref{sec:cpista}.
The PISTA curve finally corresponds to the parallel, circulant-unaware, ISTA version we proposed in~\cite{pcs2013}.

\begin{figure}[t]
  \begin{center}
   \rotatebox{270}{
	\includegraphics[width=\gnuplotFigResizeFactor\columnwidth]{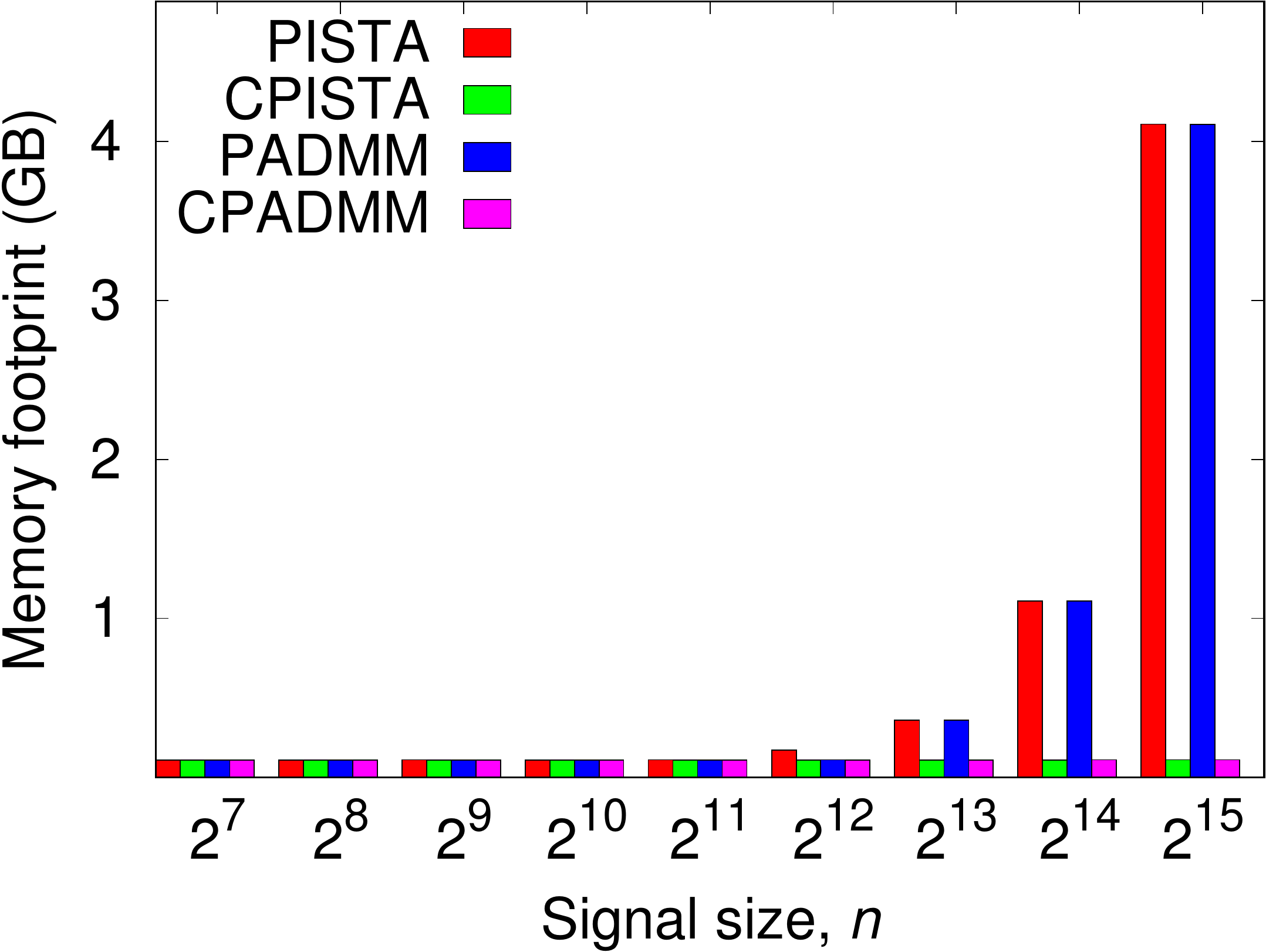}
	}
  \end{center}
  \caption{GPU memory footprint (lower is better): PADMM and PISTA footprint grows quadratically with the signal size $n$, whereas CPADMM and CPISTA footprint grows linearly only.}
  \label{fig:footprint}
\end{figure}

To start with, Fig.~\ref{fig:footprint} compares the algorithms memory footprint as a function of the signal size $n$.
We recall that in our implementations all numbers are represented over 4-bytes floats and nvidia-smi reports that about 112 MB of GPU memory are allocated during the initialization stage of each algorithm.
The experiment verifies that PISTA and PADMM footprint increases quadratically with $n$.
In fact, PADMM footprint is driven mainly by matrix $\mathbf{B}$ footprint, which increases as $n^2$.
Similarly, PISTA stores in GPU memory the $m~\times~n$ sensing matrix $\mathbf{A}$ and its transpose $\mathbf{A}^{\mathsf{T}}$, so its footprint grows nearly as $n^2$.
For example, recovering a VGA image (640$\times$480 pixels, i.e. $n \simeq 2^{18}$) with ISTA or ADMM would require more than the 16 GigaBytes of GPU memory available on our Titan X board.
Conversely, CPADMM and CPISTA footprint grows only linearly with $n$ (namely, it is equal to $4n$ for CPISTA and $10n$ for CPADMM in our implementations).
For example, recovering a VGA image with CPADMM or CPISTA would takes only about 5  and 12 megabytes of GPU memory respectively.
\\
Concluding, we showed that circulant matrices allow CPADMM and CPISTA to recover large signals that standard PADMM and ISTA would simply be unable to deal with due to their inherent memory requirements.

\begin{figure}[t]
  \begin{center}
   \rotatebox{270}{
  \includegraphics[width=\gnuplotFigResizeFactor\columnwidth]{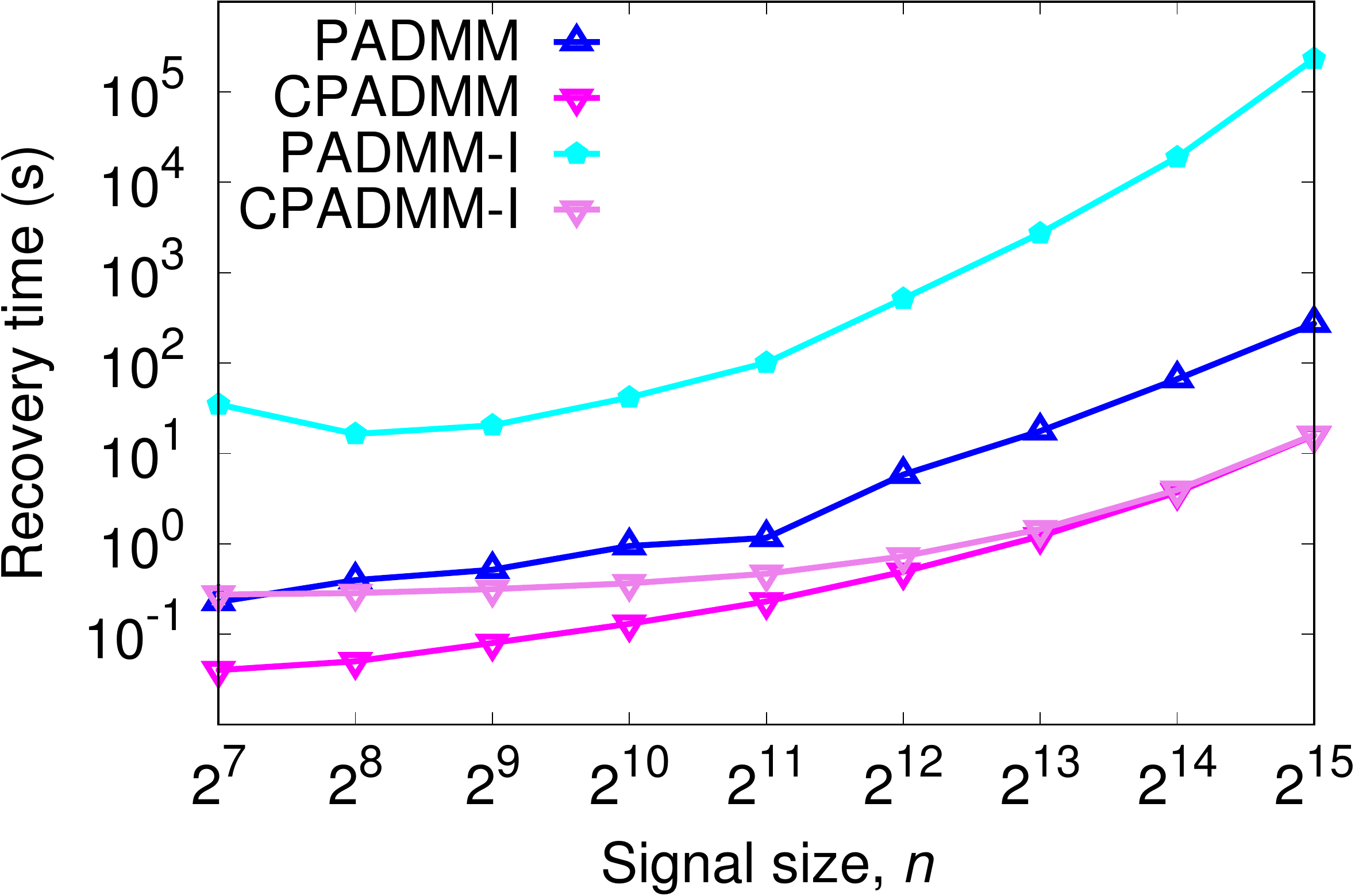}
  }
  \end{center}
  \caption{ADMM Recovery time: circulant matrices reduce ADMM recovery time (PADMM vs. CPADMM). The speedup is even larger if the initial inversion time is kept into account (PADMM-I vs. CPADMM-I).}
  \label{fig:recovery_time_admm}
\end{figure}

Next, Fig.~\ref{fig:recovery_time_admm} shows ADMM recovery time as a function of the sampled signal size $n$.
First, we measure the recovery time without accounting for the initial inversions time (PADMM and CPADMM curves).
CPADMM recovers the signal on the GPU about 10 times faster than PADMM.
CPADMM and ADMM boast the same degree of GPU parallelism, so CPADMM performance gain is due to its efficient memory representation of circulant matrices.
Efficient access to circulant matrices in GPU memory reduces the number of accesses to unique memory locations, improving the GPU caching mechanism efficiency and reducing recovery time.
\\
Next, the PADMM-I and CPADMM-I curves account also for the initial inversion time.
We recall that in both cases inversions are performed on the CPU.
However, PADMM performs a complete inversion whereas CPADMM relies on a fast FFT-based inversion.
The large gap between PADMM and PADMM-I curves shows that complete matrix inversion impacts heavily on the overall recovery time (the recovery time increase is over a tenfold factor).
Conversely, the gap between CPADMM and CPADMM-I curves shows that the FFT-based matrix inversion complexity is negligible (less than 1 s even for large $n$ values).
A comparison between CPADMM-I and PADMM-I also shows that CPADMM overall speedup over PADMM ranges between a 10-fold and an 1000-fold factor depending on $n$.
\\
Concluding, CPADMM largely outperforms a reference ADMM thanks to efficient inversion and representation in memory of circulant matrices.

\begin{figure}[b]
  \begin{center}
   \rotatebox{270}{
  \includegraphics[width=\gnuplotFigResizeFactor\columnwidth]{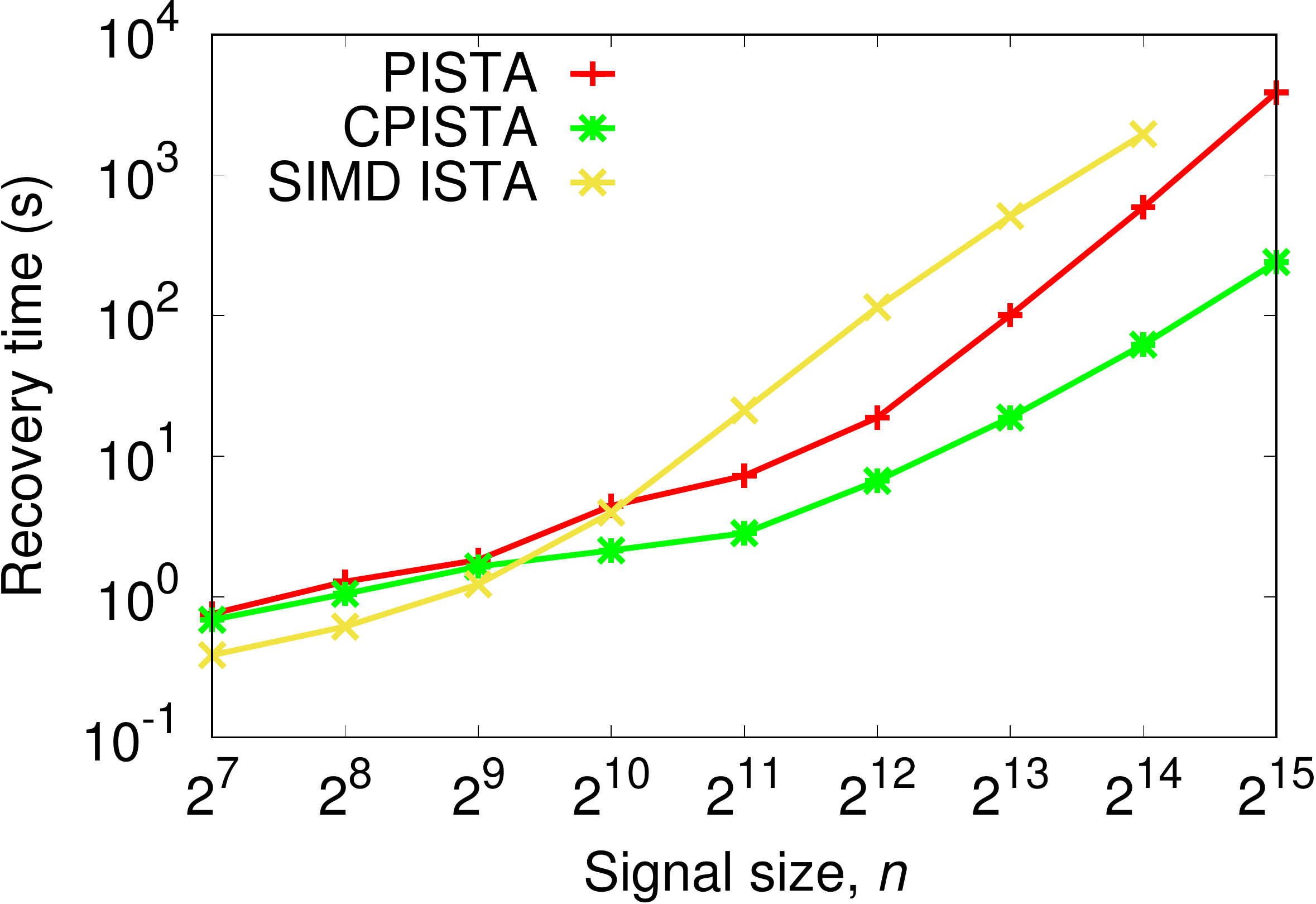}
  }
  \end{center}
  \caption{ISTA Recovery time: circulant matrices enable a tenfold reduction of recovery time for ISTA exploiting GPU core-level parallelism.}
  \label{fig:recovery_time_ista}
\end{figure}

Similarly, Fig.~\ref{fig:recovery_time_ista} shows ISTA recovery time as a function of $n$.
As a further reference, the figure includes \emph{SIMD-ISTA}, a single-core, CPU-only, ISTA implementation that exploits the SSE SIMD instructions of the x64 architecture (i.e., it offers data-level parallelism on the CPU).
PISTA recovery time grows more than linearly with $n$: as the footprint increases, the GPU caching mechanism becomes less effective and memory latency increases.
As $n$ increases, the gap between PISTA and SIMD-ISTA becomes thinner, i.e. the GPU effective level of parallelism drops when memory latency increases to the point where a GPU-based version is only marginally better than a CPU-based reference.
Conversely, CPISTA recovery time grows almost linearly with $n$ due to the reduced memory footprint that enables efficient use of GPU caching mechanism.
Overall, CPISTA logs a 10-fold recovery time reduction over PISTA for large $n$ values.
\\
Concluding, the experiments show that our CPISTA version significantly speedup the ISTA algorithm increase thanks to efficient representation of circulant matrices in GPU memory.

\begin{figure}[h!]
    \centering
     \subfigure{
      \rotatebox{270}{
        \includegraphics[width=1.8in]{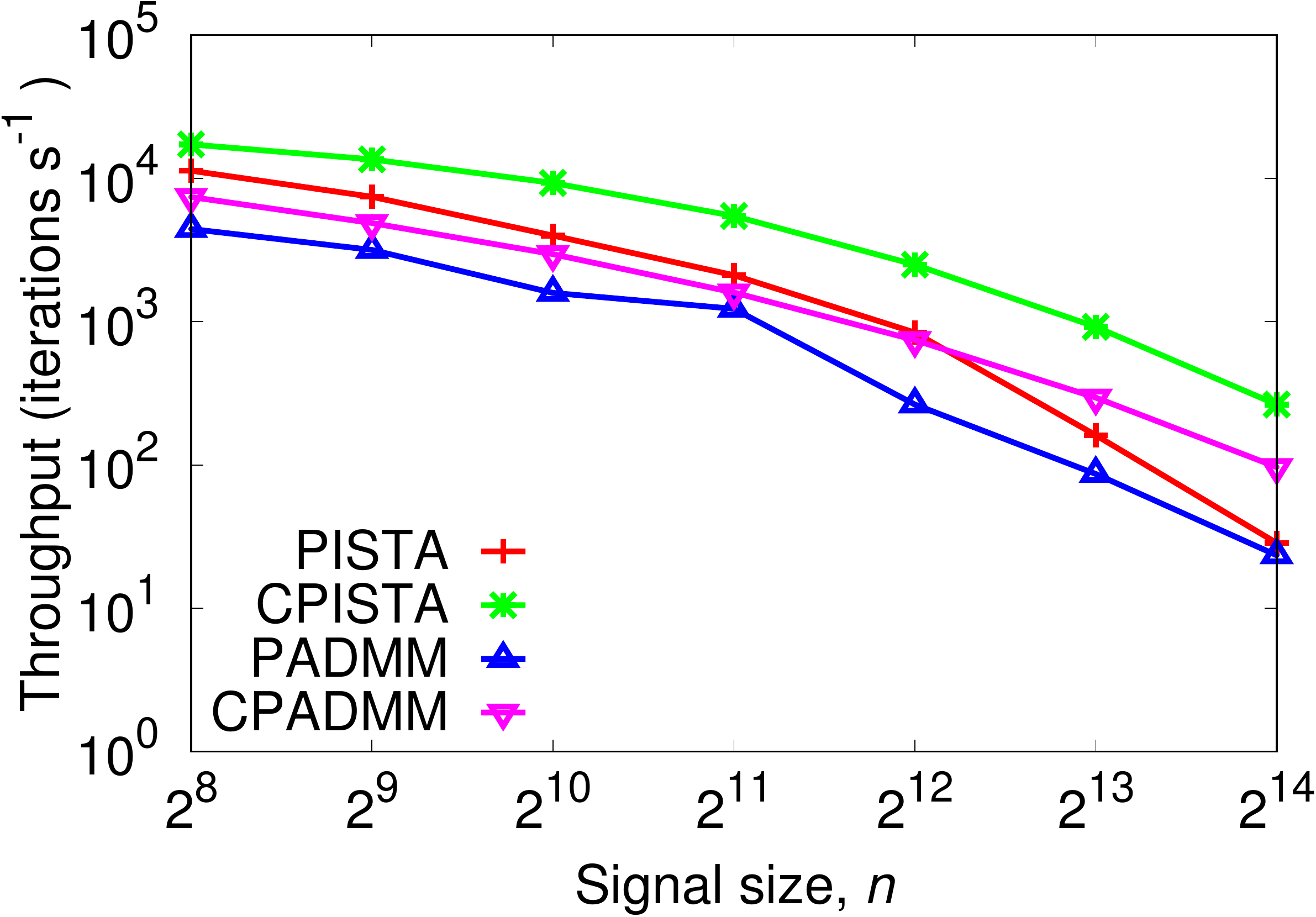}
      }
      \label{fig:dista_time_gpu}
    }
    \subfigure{
      \rotatebox{270}{
        \includegraphics[width=1.8in]{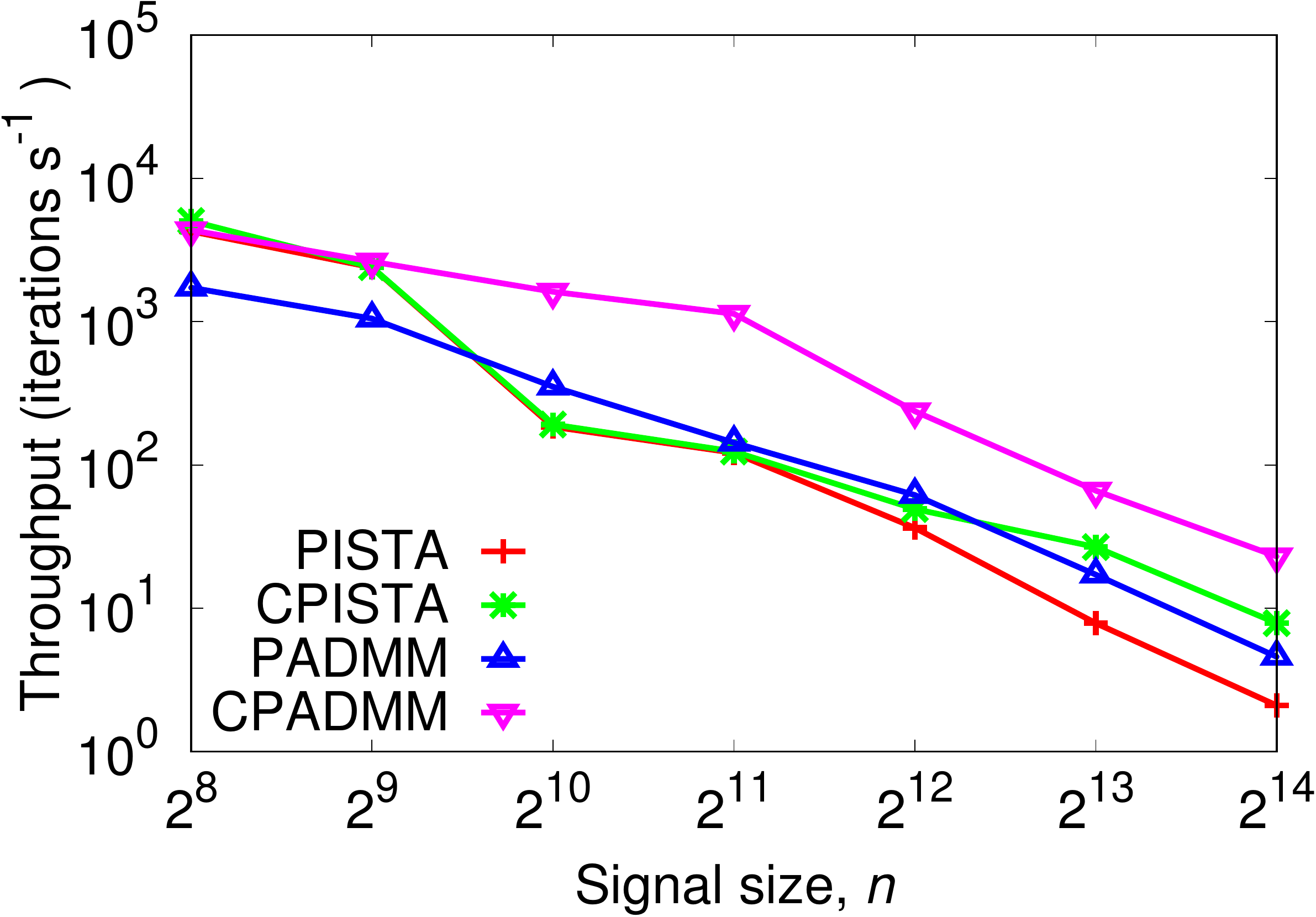}
      }
      \label{fig:dista_time_cpu}
    }
    \caption{Algorithmic throughput (higher is better) measured as number of iterations per second as a function of the signal size $n$ to recover over two different types of core-level parallel architectures. Left graph: NVIDIA GPU with 4096 cores. Right graph: Intel x64 CPU with 64 cores.}
    \label{fig:dista_time}
\end{figure}

Fig.~\ref{fig:dista_time} shows the ISTA and ADMM algorithmic throughput expressed as number of iterations per second as a function of the signal size $n$.
The left graph shows the throughput of our algorithms as measured on our NVIDIA TitanX GPU.
The graph provides an insight on why CPADMM and CPISTA iterate faster than their PADMM and PISTA counterparts as shown in Fig.\ref{fig:recovery_time_admm}.
Because of the fewer memory accesses, CPADMM and CPISTA complete each iteration in less time than their counterparts PADMM and PISTA, delivering higher throughput.
\\
As our OpenCL algorithmic implementation enables us to experiment over different devices simply by switching to the appropriate backend, in the right graph we benchmarked our algorithms over a quad-socket Intel E5-2690 server (eight x64 cores per socket, for a total of 32 x64 cores).
The graph shows that the use of circulant matrices improves the ISTA and ADMM performance also over x64 CPUs, albeit the CPUs lower degree of parallelism yields lower overall performance than GPUs.

\begin{figure}[b!]
  \begin{center}
   \rotatebox{270}{
  \includegraphics[width=\gnuplotFigResizeFactor\columnwidth]{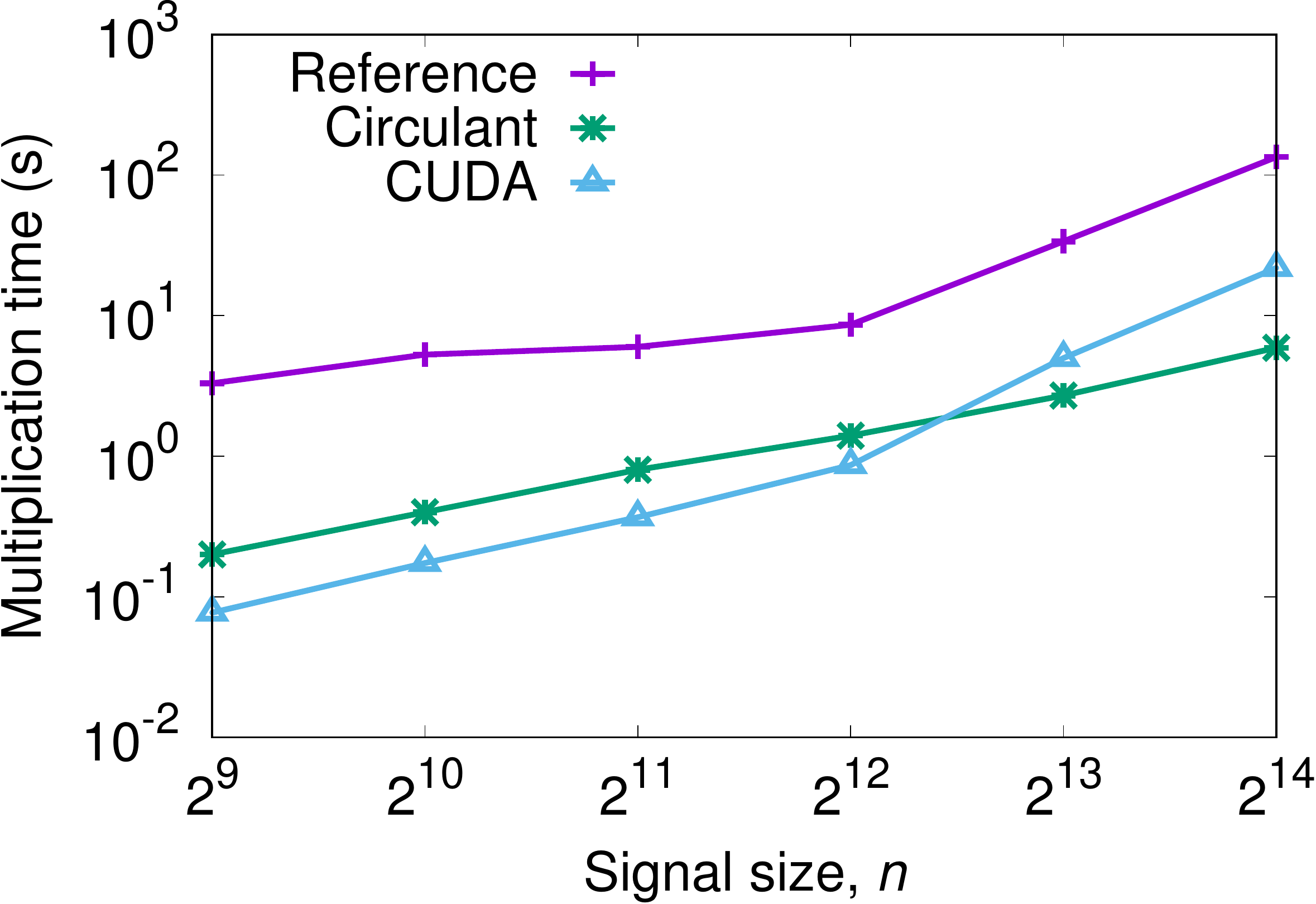}
  }
  \end{center}
  \caption{Matrix-vector multiplication time as a function of the matrix edge size $n$: as $n$ increases, a multiplication scheme leveraging the properties of circulant matrices outperforms an highly-optimized scheme that does not leverages such properties.}
  \label{fig:cuda}
\end{figure}

To acquire a further insight in the role of circulant matrices role in improving the performance of our algorithms, we experimented with CUDA.
CUDA is NVIDIA framework for GPU programming and includes highly-optimized libraries for matrix operations such as the \emph{cuBLAS} library.
Namely we evaluate the performance of matrix-vector multiplication $\mathbf{A}~\bm{b}$, which are the major sources of ISTA and ADMM complexity where $\mathbf{A}$ is an $n \times n$ matrix of single precision floating point numbers.
We consider three different schemes.
The Reference scheme is such that for each of the $n$ rows of $\mathbf{A}$, one OpenCL thread is issued. The thread multiplies each element of the i-th row of $\mathbf{A}$ by the i-th element of vector $\bm{b}$ over a for loop, accumulates the partial results in a private accumulator and eventually stores the result back in global memory.
In the \emph{Circulant} scheme, $\mathbf{A}$ is assumed to be circulant and therefore each \emph{i}-th thread accesses only the first $\mathbf{A}$ row starting from the \emph{i}-th column and wrapping around the matrix boundary.
Finally, in the CUDA scheme the $\mathbf{A}~\bm{b}$ multiplication is executed leveraging the \emph{cublasSgemm()} function provided by \emph{cuBLAS}.
We point out that the Reference and CUDA schemes access the same number of unique addresses in global memory during the fetch stage, which is equal to $2n+n$.
Conversely, the Circulant scheme accesses only $2n$ unique locations in global memory.
Fig.~\ref{fig:cuda} shows the average time required by each matrix multiplication for each considered scheme.
The CUDA scheme performs better than Reference for any $n$ and performs better than Circulant for small $n$ values.
However, as $n$ increases, the Circulant scheme performs increasingly better than CUDA due to the fewer accesses to global memory.
Concluding, a straightforward matrix-vector multiplication scheme that leverages the properties of circulant matrices outperforms highly efficient matrix-vector implementation as the algorithm working set increases in reason to the fewer accesses to the global memory.

\begin{figure}[b!]
  \begin{center}
   \rotatebox{270}{
	\includegraphics[width=\gnuplotFigResizeFactor\columnwidth]{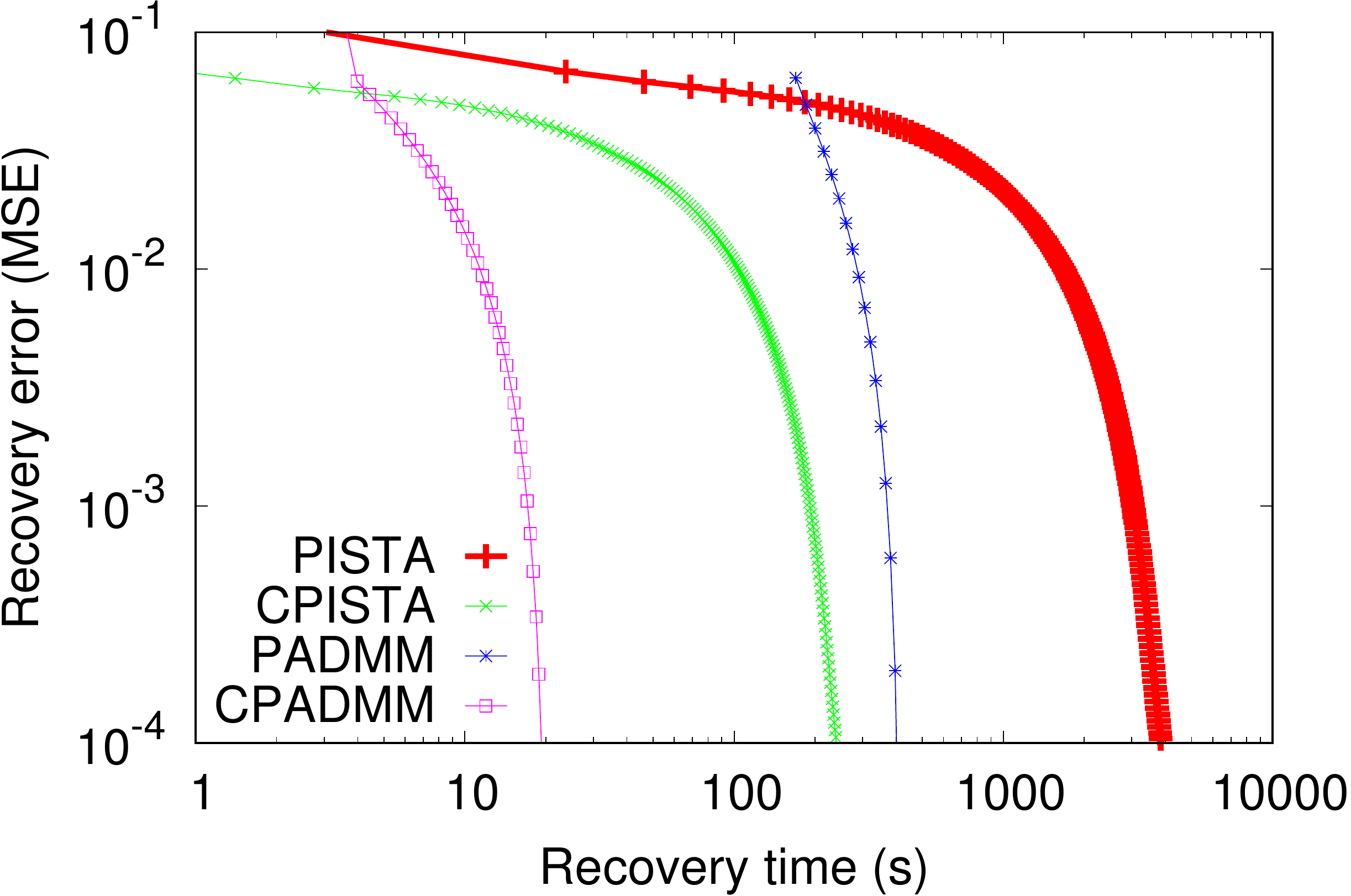}
	}
  \end{center}
  \caption{Recovery error (MSE) over time: depending on the considered target error, ISTA may converge to the target error faster than ADMM.}
  \label{fig:graph_ista_vs_admm}
\end{figure}

Finally, Fig.~\ref{fig:graph_ista_vs_admm} shows the recovery error of our algorithms over time for a fixed signal size $n = 2^{15}$.
Notice that for PADMM and CPADMM, we account also for the initial inversion time.
CPADMM and CPISTA error decreases faster than PADMM and PISTA for the reasons explained above.
ADMM is such that the recovery error remains constant until the initial inversions is completed, as the actual signal recovery begins only afterwards.
Conversely, CPISTA error starts to decrease immediately, and for targets MSE above 0.05, CPISTA converges faster than CPADMM in this experiment.
\\
This experiment suggest that, depending on the considered target error, CPISTA may recover the signal (i.e., achieve the target error rate) faster than CPADMM, albeit the latter converges faster to very small target error values.

\section{An Application: Compressed Image Deblurring}
\label{sec:deblurring}

In this section we demonstrate the CPADMM algorithm in deblurring a compressed astronomical image.
Astronomical images are often affected by blurring due to the atmospheric turbulence \cite{ami13}, thus deblurring is required.
Generally speaking, blur can be modeled as the convolution between a blurring filter $\mathbf{B}$ and image $\bm{x}$: the result is the observed blurred image.
Deblurring the image may be achieved by inverse-recovering the signal $\bm{x}$ convolved with filter $\mathbf{B}$, but such practice is not recommended for it may amplify the noise.

Several convolutional filters can be however represented by circulant matrices. Considering a filter of order $L$, the blurring matrix $\mathbf{B}$ is obtained by right-circulating a vector with the first $L$ components equal to $\frac{1}{L}$ and zero otherwise.
Let the blurring matrix $\mathbf{B}$ be a $n \times n$ circulant matrix: we can multiply it by a circulant $m \times n$ sensing matrix $\mathbf{C} (m < n)$, and obtain a circulant $m \times n$ matrix $\mathbf{A}=\mathbf{C}\mathbf{B}$.
If image $\bm{x}$ is sparse as many astronomical images are, the problem of recovering $\bm{x}$ given $\bm{y}=\mathbf{A}\bm{x}$ can be recast as a CS problem where $\mathbf{A}$ is a circulant sensing matrix.
In other words, we can jointly sparse-sample and deblur the observed signal to perform a \emph{compressed deblurring},
for example solving the LASSO problem.

\begin{figure*}[t]
    \centering
    \subfigure[(\textit{a})~Original image]
    {
        \includegraphics[width=2.1in]{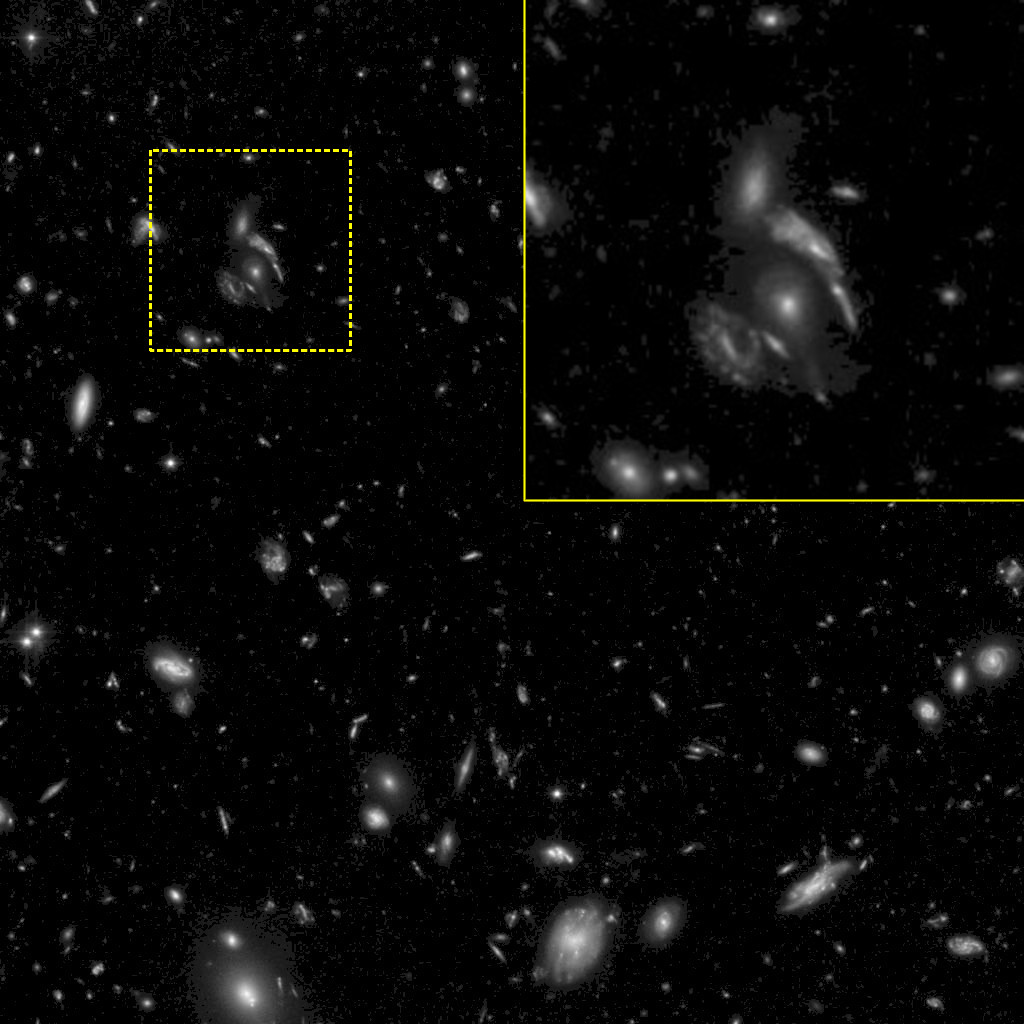}
        \label{fig:deblurring_a}
    }
    \subfigure[(\textit{b})~Blurred image (detail)]
    {
        \includegraphics[width=2.1in]{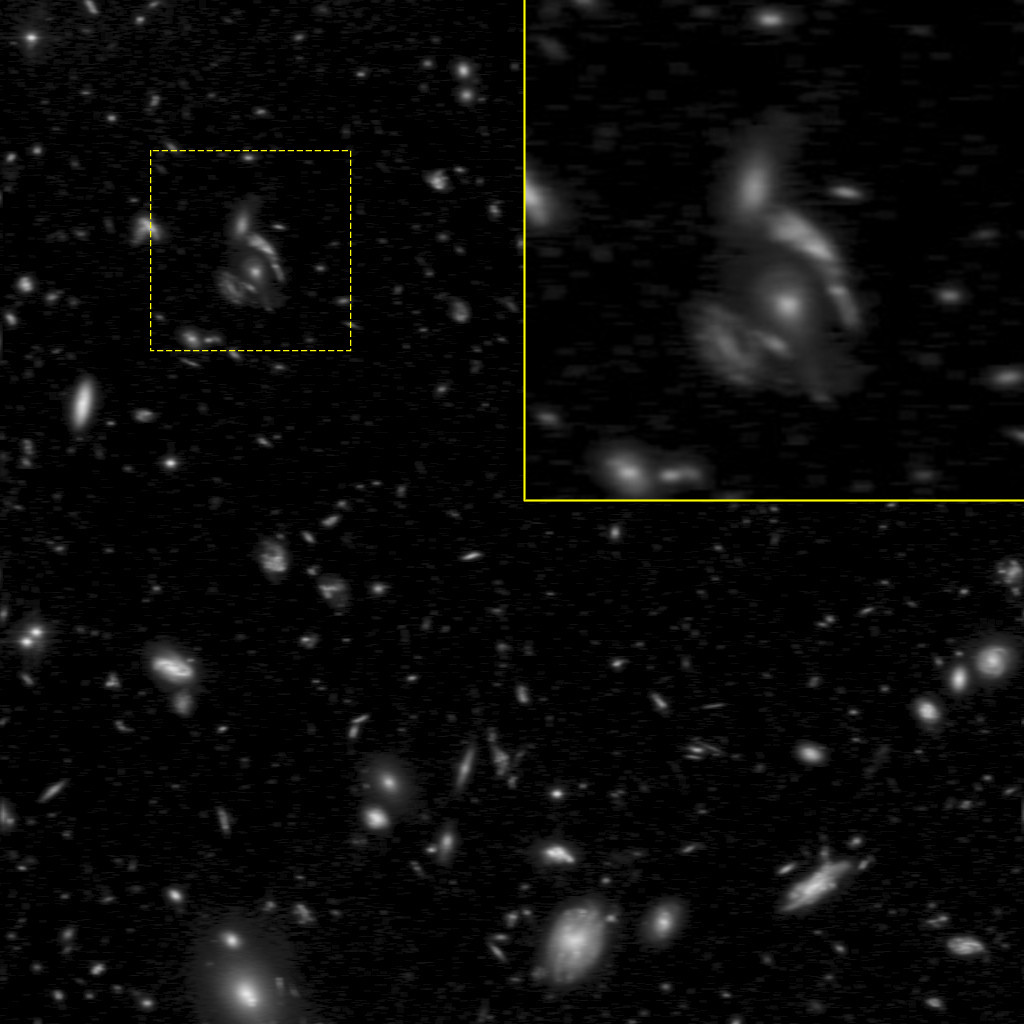}
        \label{fig:deblurring_b}
    }
    \subfigure[(\textit{c})~Recovered image (detail)]
    {
        \includegraphics[width=2.1in]{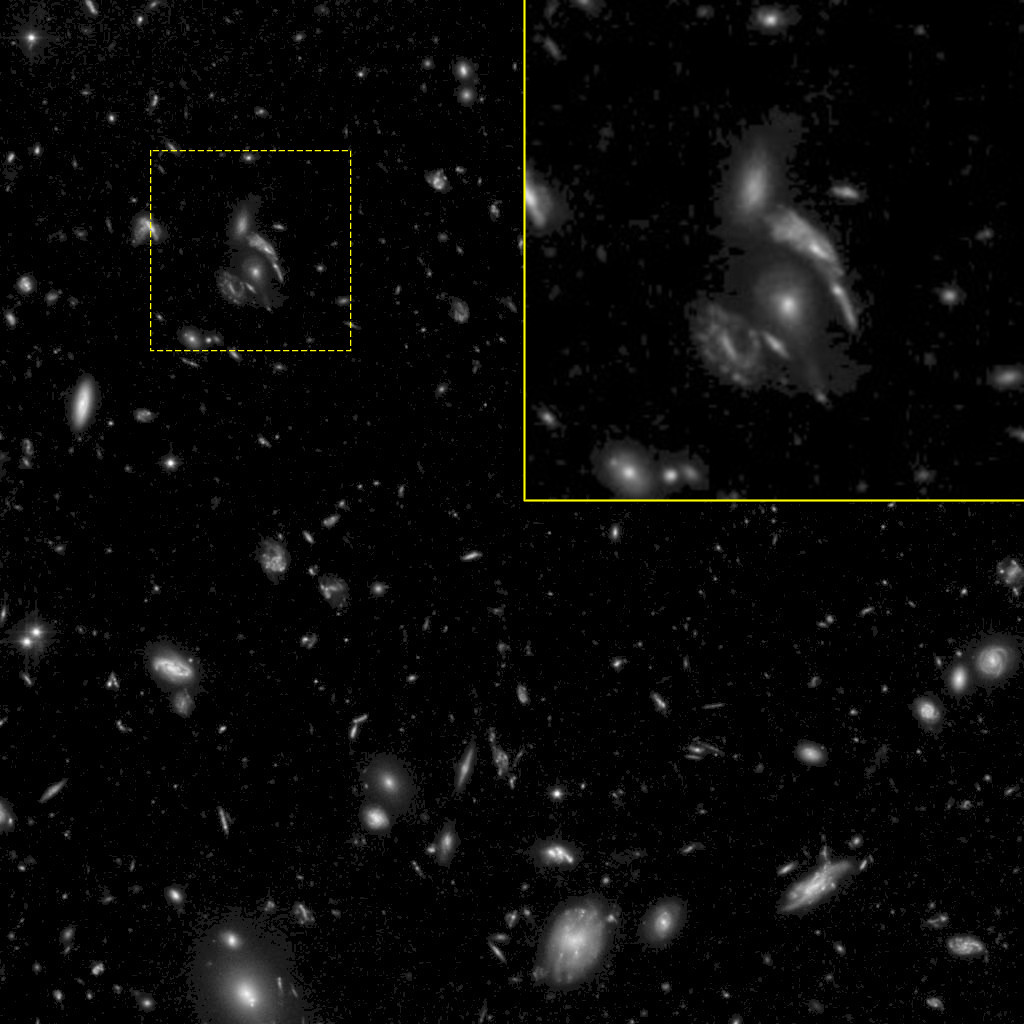}
        \label{fig:deblurring_c}
    }
    \subfigure[(\textit{d})~Absolute recovery error]
    {
        \includegraphics[width=2.1in]{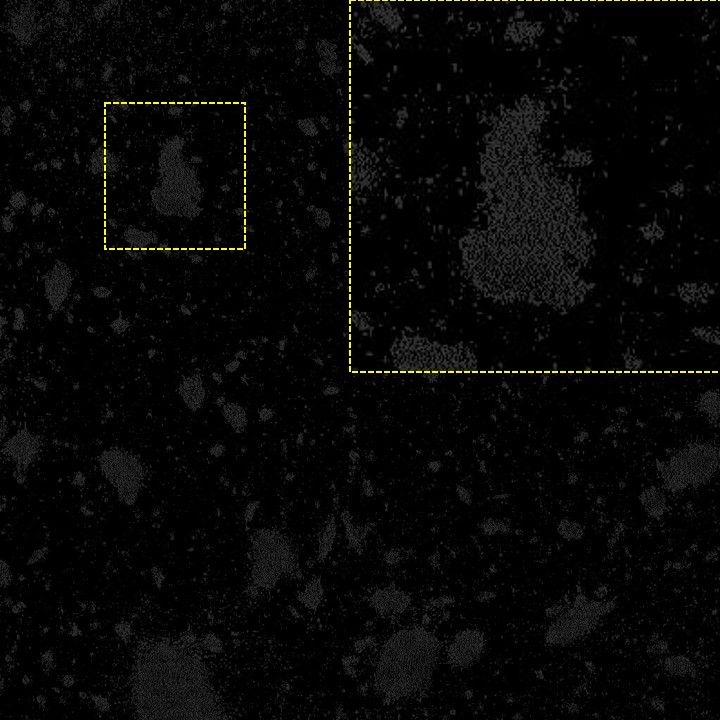}
        \label{fig:deblurring_d}
    }
    \caption{Compressed deblurring of a 1024x1024 astronomical image ($L$=5). Original image (\textit{a}), blurred image (\textit{b}), recovered deblurred image (\textit{c}) and recovery error (\textit{d}). The top-right corner zooms over the dotted region in the full-scale pictures: the deblurred image is very similar to the original image also at a close inspection.}
	\label{fig:deblurring}
\end{figure*}

Fig.~\ref{fig:deblurring_a}~(\textit{a}) shows the $1024 \times 1024$ pixel image used for the experiment: it depicts \emph{Abell 2744} (also known as \emph{Pandora's box}) cluster as captured by the Hubble telescope.
The image is represented as vector $\bm{x}$ of size $n \simeq 10^6$ and is naturally sparse, as many pixel are almost black,
with a sparsity of about $10\%$ of the signal size.
We apply a one-dimensional blurring filter $\mathbf{B}$ of order $L=5$ obtaining a blurred image representing the same scene as observed on Earth accounting for atmospheric turbulence.
Fig.~\ref{fig:deblurring_b}~(\textit{b}) shows the same image as acquired from a hypothetical observer on Earth: the blurring effect is visible with the naked eye.
We sparse-sample the blurred image with $m = \frac{n}{2}$ and process the samples with the CPADMM algorithm ($\alpha = 10^{-2}$).
Fig.~\ref{fig:deblurring_c}~(\textit{c}) shows the image as recovered via CPADMM after about 300 iterations and 3 hours of processing (the original-recovered MSE is in the order of $10^{-2}$, and the corresponding normalized MSE is in the order of $10^{-4}$).
A comparison with Fig.~\ref{fig:deblurring_a}~(\textit{a}) reveals that the recovered image is more crisp than the acquired image thanks to the removal of the blurring artifacts.
Fig.~\ref{fig:deblurring_d}~(\textit{d}) additionally shows the pixel-wise absolute error between the recovered image and the original image normalized with respect to the original image mean intensity value (its mean value is equal to 0.0157).
A comparison of the original and the recovered images shows that no portions of the recovered image are significantly degraded, and thus the recovery process is satisfying from a perceptual perspective.
Finally, the corresponding memory footprint is equal to 163 Mbytes, that is our CPADMM only required 163 Mbytes of GPU memory to recover this image.Such figure is well within the capabilities of modern desktop-class GPUs, which enables to recover of large imaging signals even with consumer-grade electronics.

\renewcommand{\thesubfigure}{}

\section{Conclusions and Lessons Learned}
\label{sec:conclusions}

We described CPISTA and CPADMM, two parallel algorithms for sparse image recovery that leverage the properties of circulant matrices to speedup the signal recovery process.
Namely, the properties of circulant matrices are key to reduce the number of sequential accesses to global memory, increasing the algorithms practically attainable level of parallelism.
Our algorithms are implemented in OpenCL language, enabling straightforward deployment on GPUs and multicore CPUs.
Our experiments showed up to tenfold gains over reference algorithms that do not leverage circulant matrices properties.
Finally, we practically demonstrated deblurring a megapixel ($10^6$ pixel) image on a desktop GPU, showing that circulant matrices enable recovering even large signal with limited hardware requirements.



\bibliographystyle{plain}
\bibliography{main}

\end{document}